# Experimental Methods, Health Indicators, and Diagnostic Strategies for Retired Lithium-ion Batteries: A Comprehensive Review

**Authors**

Song Zhang[1], Ruohan Guo[3], Xiaohua Ge[1], Perter Mahon[2], Weixiang Shen[1,*]

**Affiliations**

1. School of Engineering, Swinburne University of Technology, Hawthorn VIC 3122, Australia

2. School of Science, Computing and Emerging Technologies, Swinburne University of Technology, Hawthorn VIC 3122, Australia

3. Department of Electrical and Electronic Engineering, Research Centre for Grid Modernisation, The Hong Kong Polytechnic University, Kowloon, Hong Kong, China

*Corresponding author: Weixiang Shen, Email: wshen@swin.edu.au

Table of Contents

## Abstract

Reliable health assessment of retired lithium-ion batteries is essential for safe and economically viable second-life deployment, yet remains difficult due to sparse measurements, incomplete historical records, heterogeneous chemistries, and limited or noisy battery health labels. Conventional laboratory diagnostics, such as full charge–discharge cycling, pulse tests, Electrochemical Impedance Spectroscopy (EIS) measurements, and thermal characterization, provide accurate degradation information but are too time-consuming, equipment-intensive, or condition-sensitive to be applied at scale during retirement-stage sorting, leaving real-world datasets fragmented and inconsistent. This review synthesizes recent advances that address these constraints through physical health indicators, experiment testing methods, data-generation and augmentation techniques, and a spectrum of learning-based modeling routes spanning supervised, semi-supervised, weakly supervised, and unsupervised paradigms. We highlight how minimal-test features, synthetic data, domain-invariant representations, and uncertainty-aware prediction enable robust inference under limited or approximate labels and across mixed chemistries and operating histories. A comparative evaluation further reveals trade-offs in accuracy, interpretability, scalability, and computational burden. Looking forward, progress toward physically constrained generative models, cross-chemistry generalization, calibrated uncertainty estimation, and standardized benchmarks will be crucial for building reliable, scalable, and deployment-ready health prediction tools tailored to the realities of retired-battery applications.

# 1 Introduction

The rapid electrification of transportation and the widespread deployment of lithium-ion batteries (LIBs) have led to an unprecedented growth in retired battery volume worldwide. As electric vehicles reach mid-life and fleets undergo replacement cycles, millions of end-of-life or retired LIBs (R-LIBs) are expected to enter the secondary market annually over the next decade [1]. These retired cells typically retain 70–80% of their initial capacity and therefore present a valuable opportunity for repurposing in low-to-medium-power stationary storage, renewable energy buffering, community microgrids, and distributed energy resources [2]. Leveraging these partially degraded batteries is widely recognized as a key pathway toward reducing lifecycle carbon emissions, lowering the cost of energy storage, and promoting a more circular battery economy. However, the successful adoption of second-life systems depends critically on the ability to assess health, safety, and performance in a reliable, scalable, and economically viable manner.

Despite their remaining value, retired batteries exhibit highly heterogeneous aging behaviors that make their health state intrinsically difficult to determine. Cells are retired from diverse applications, undergo vastly different load profiles, and are exposed to varying temperatures, depths of discharge, calendar storage conditions, and charging protocols over their previous use phase [3]. This diversity results in broad distributions of capacity fade, internal resistance, rate capability, and thermal stability across batteries sourced from the same vehicle pack or logistics batch. In addition, the operational history of retired batteries is often unavailable or incomplete, which prevents conventional model-based diagnosis from being applied with confidence. As a result, even batteries with similar nameplate specifications can present substantially different degradation modes, ranging from loss of lithium inventory and active-material depletion to impedance rise, gas generation, and structural instabilities. These uncertainties make reliable classification and sorting extremely challenging and represent one of the largest technical barriers to deploying second-life systems at industrial scale.

Current research addressing these challenges remains fragmented across experimental testing, signal-based health indicators, and data-driven prediction approaches. Experimental studies have proposed a wide range of diagnostic procedures, including capacity tests, direct-current internal resistance (DCIR), electrochemical impedance spectroscopy (EIS), pulse-power characterizations, and open-circuit voltage (OCV) relaxation analysis, to quantify the degradation state of R-LIBs [4]. Meanwhile, the signal-processing community has introduced voltage-based, derivative-based, impedance-derived, thermal, acoustic, and visual indicators to interpret the physical mechanisms underlying aging. Parallel to these developments, modelling research has advanced machine learning and deep learning frameworks for end-of-life prediction, state-of-health estimation, and multi-modal feature extraction, typically based on high-quality datasets collected under controlled laboratory conditions [5]. However, these research threads often evolve independently, using inconsistent protocols, incompatible data formats, and different modelling assumptions. This fragmentation makes it difficult to compare results across studies and, more importantly, prevents the formation of a unified methodology suited for large-scale, real-world retired-battery deployment.

Given these limitations, a comprehensive and integrative review that bridges experimental diagnosis and data-driven modelling is urgently needed. Existing reviews have focused on either battery recycling routes, second-life techno-economic assessment, or standard health-estimation methods for in-vehicle batteries, but relatively few studies systematically examine the specific challenges of retired batteries. For example, their heterogeneous, incomplete, and often noisy health signals; their non-standardized testing conditions; and their stronger safety constraints [6]. This review aims to consolidate knowledge across diagnostic techniques, health indicators, signal interpretation, and predictive modelling, providing a coherent framework for understanding how raw measurements can be transformed into actionable health knowledge for R-LIB screening, sorting, and repurposing. In particular, we highlight how different classes of health indicators capture complementary degradation mechanisms, how missing or irregular data can be reconstructed or generated, and how modern machine-learning models can be adapted for low-information, high-variability retired-battery scenarios.

The remainder of this review is organized as follows. Section 2 outlines the practical challenges associated with retired-battery utilization, including the intrinsic uncertainty of health states, inconsistent assessment protocols, and safety-related constraints. Section 3 provides a comprehensive overview of health assessment methods for retired lithium-ion batteries, covering preliminary screening routes, physically interpretable health indicators, and diagnostic experiments that translate raw signals into quantitative grading criteria. Section 4 focuses on health estimation and prediction under data scarcity, examining the role of synthetic data generation, supervised and semi-supervised learning, weakly supervised approaches, and fully unsupervised methods. This section also compares representative modelling routes and highlights emerging opportunities for robust prediction in low-information scenarios. Section 5 concludes the review with a summary of key insights and implications for future second-life deployment.

## 2 Challenges in Retired Battery Utilization

Repurposing retired lithium-ion batteries offers substantial economic and environmental benefits, but their practical deployment is constrained by a set of inherent challenges arising from heterogeneous aging behaviors, inconsistent assessment practices, and heightened safety risks. These issues are fundamentally different from those encountered in in-vehicle battery management, as retired cells enter the second-life stream with highly diverse degradation histories and uncertain safety margins. With global EV adoption accelerating, the volume of R-LIBs is projected to reach 100–120 GWh annually by 2030, equivalent to today's total yearly battery production, introducing more than one million EV packs per year into the repurposing or disposal pipeline [7], [8], [9]. This industrial-scale influx underscores the urgency of developing reliable, fast, and standardized health assessment frameworks [1], [10].

A first major challenge stems from the intrinsic uncertainty of health states in retired batteries. R-LIBs originate from heterogeneous operational environments with unknown duty cycles, varying depth-of-discharge patterns, inconsistent thermal exposure, and irregular storage or abuse conditions [11], [12]. Even cells of the same chemistry or manufacturer can exhibit markedly different residual capacity, impedance rise, rate capability, or thermal stability. Most EV batteries are retired at around 70–80% of their nominal capacity, yet capacity-based retirement alone does not guarantee structural or electrochemical integrity [13]; latent defects such as micro-cracks, localized lithium plating, or separator degradation may remain undetected but significantly elevate safety risks during reuse [14]. Reported incidents of thermal runaway in repurposed cells illustrate how subtle internal degradation, missed by coarse screening, can propagate rapidly under load. At a more fundamental level, the underlying degradation mechanisms include SEI thickening [15], loss of active lithium [16], particle cracking [17], electrolyte decomposition[18], and gas generation, none of which can be directly accessed at scale without destructive teardown or advanced laboratory characterization [19], [20]. As a result, health estimation must be inferred indirectly from macro-scale observables, complicating diagnosis and amplifying uncertainty.

This challenge is further compounded by pronounced heterogeneity in physical and electrochemical performance. Even within the same pack, capacity spread can reach 20–30%, while internal resistance may vary by a factor of two to three due to variations in temperature cycling, manufacturing tolerances, and nonlinear degradation dynamics. Differences in cathode chemistry (e.g., LFP, NCM, NCA), form factors (pouch, cylindrical, prismatic), and production vintage introduce additional variability that fundamentally reshapes the mapping between observable indicators and true degradation state [4]. Such diversity necessitates more sophisticated, statistically informed sorting strategies and contributes to inconsistent grading outcomes across facilities.

A second critical challenge arises from the lack of standardized testing and assessment protocols. Current industrial and research practices employ widely differing current rates, voltage windows, ambient conditions, and rest intervals for capacity tests [21], [22]. Internal resistance assessment suffers from similar inconsistency: DCIR, pulse-power measurement, hybrid pulse power characterization (HPPC), and EIS are conducted using non-uniform current amplitudes, frequency ranges, durations, and instrumentation [23], [24]. These variations severely limit cross-laboratory comparability, hinder data fusion across sources, and obstruct the development of universal sorting criteria. Furthermore, full capacity tests require hours per cell and are prohibitively slow for industrial throughput, while high-fidelity EIS or thermal characterization demands specialized equipment rarely available in large-scale repurposing lines. The absence of harmonized and scalable protocols ultimately results in fragmented datasets and reduced confidence in second-life applications.

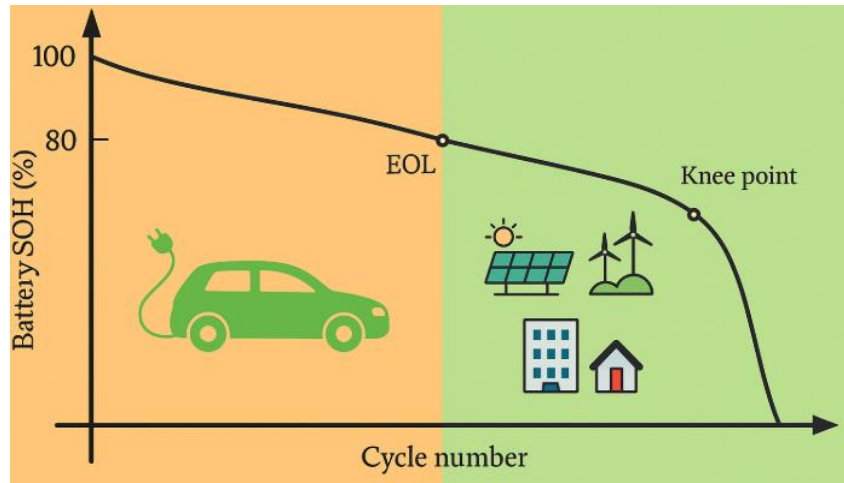

Fig.1 Conceptual aging trajectory of lithium-ion batteries (adapted from [25])

Beyond uncertainty and protocol inconsistency, safety considerations impose significant additional constraints. Aging-driven phenomena such as impedance growth, lithium plating, micro-short formation, gas evolution, and structural weakening increase the probability of thermal instability during reuse, particularly under high-current or elevated-temperature operation [26]. Retired cells are prone to localized hotspots, current inhomogeneity, and accelerated degradation when placed into new module configurations that differ from their original pack environments. Even cells that appear electrically functional may harbor latent internal failures, such as separator thinning or partial internal short circuits, that evade basic electrical tests but substantially elevate the risk of failure propagation. Safety risks are further exacerbated when cells with diverse chemistries, aging levels, or manufacturers are mixed within the same second-life system, complicating thermal management and reducing overall stability.

Taken together, these challenges highlight the need for accurate, rapid, and scalable approaches to retired-battery health assessment [27]. Health uncertainty necessitates robust indicators capable of capturing the diverse signatures of degradation across electrochemical, impedance, and thermal domains. Protocol inconsistency emphasizes the importance of harmonized diagnostic pathways and physically interpretable metrics that remain comparable across testing conditions. Safety concerns call for multi-modal diagnostic tools and predictive models that detect precursors of failure beyond traditional thresholds. These considerations motivate the detailed examination of health indicators, diagnostic pathways, and predictive modelling strategies presented in the subsequent sections.

## 3  Health Assessment of Retired Lithium-ion Batteries
## 3.1  Preliminary Screening and Diagnostic Pathways

Before a R-LIB can be repurposed for second-life use, its condition must be verified through a structured and reliable screening workflow. The purpose of preliminary screening is not only to estimate the residual health of each unit but also to identify and remove cells exhibiting structural or electrical anomalies that could compromise safety in downstream applications. Unlike the controlled environment of in-vehicle battery management, second-life facilities must evaluate incoming cells rapidly and consistently, often in the absence of detailed historical information. Consequently, screening procedures are designed to balance diagnostic accuracy with throughput, ensuring that only units with acceptable mechanical integrity and electrical behavior proceed to more detailed assessment stages.

Industrial practice typically adopts a hierarchical, top-down assessment workflow to navigate this complexity and ensure safety and efficiency in large-scale facilities. The process begins at the pack level, where parameters such as total voltage, insulation resistance, and abnormal temperature gradients are evaluated to quickly isolate units exhibiting critical faults like leakage, severe over-discharge, or thermal imbalance [28]. Modules passing this initial inspection undergo further examination for wiring integrity, terminal corrosion, and resistance mismatches within parallel strings, enabling the localization of problematic sub-assemblies without the need for full disassembly [29]. Only those modules deemed safe are subsequently opened for cell-level diagnostics, where more detailed assessments of capacity retention, impedance, and mechanical integrity are performed to determine suitability for repurposing [30]. This staged workflow not only prioritizes safety and operational efficiency but also minimizes unnecessary teardown, thereby reducing costs and processing time. In response to industrial scale demands, several pilot facilities have established automated lines integrating robotic pack disassembly, multiplexed electrical testing, and even ultrasound or thermal imaging. Cycle time remains a major challenge: a full discharge test may require hours, whereas a short pulse or IR test can sort thousands of cells per shift, albeit with some trade-off in diagnostic accuracy.

Within cell-level screening, diagnostics generally proceed from rapid exclusion toward increasingly detailed analysis. Visual and physical inspection is used first to identify cells exhibiting swelling, deformation, leakage, or casing damage, clear indicators of internal faults or compromised mechanical robustness, and thus immediate grounds for removal [31]. Units that pass this initial check are then subjected to basic electrical tests such as OCV measurement [32], [33], DCIR and short-duration charge–discharge pulses, which offer quick estimates of state of health (SOH) and power capability [34], [35]. While these rapid tests are widely deployed for their speed and simplicity, their accuracy can be strongly affected by temperature, state of charge, and test protocol variations. For cells that demonstrate promising characteristics, more advanced diagnostic methods, such as EIS, incremental capacity (IC) analysis, differential voltage (DV) analysis, and HPPC, provide richer information regarding degradation mechanisms and transient power performance [29], [36], [37], [38]. IC analysis reveals phase transitions in electrodes as distinct peaks in dQ/dV curves, which shift and diminish with aging, offering insights into loss of active material and increased kinetic resistance. Electrochemical impedance spectroscopy (EIS), increasingly paired with model-based fitting or machine learning, can decompose total impedance into physically meaningful elements; for example, growth in charge-transfer resistance ($R_{ct}$) closely tracks SEI thickening and capacity fade [39]. However, these techniques are typically reserved for a smaller subset of candidate R-LIB, given their greater time and operational requirements.

To enable high-throughput and efficient screening, recent advances have focused on accelerated diagnostic protocols. For instance, partial charging tests, such as evaluating LFP cells within a narrow voltage window, can estimate full capacity with over 50% reduction in test time [40]. Multi-index co-extraction schemes further enhance efficiency by deriving voltage, current, and temperature features from the same short testing pulse [41], [42]. In parallel, the integration of robotic handling, multiplexed hardware, and AI-assisted analytics has enabled high-throughput automated screening lines capable of evaluating thousands of R-LIBs per hour.

Following diagnostic evaluation, R-LIBs are typically classified according to a three-tier scheme based on key health metrics. Cells retaining more than approximately 80% of their nominal capacity, with low internal resistance and no visible damage, are considered suitable for demanding reuse scenarios such as mobility devices or power tools. Those with moderate SOH, generally in the 50–80% range, are allocated to less power-intensive applications including stationary energy storage and backup systems. Units exhibiting SOH below 50%, abnormal resistance growth, or clear physical defects are directed to recycling and material recovery [13], [43], [44]. While a three-tier classification scheme (SOH >80%, 50–80%, <50%) is widely adopted, exact cutoffs for end-of-life (EOL) and second-life suitability vary by jurisdiction and organization. Many automakers and standards use 80% capacity as the nominal EOL for mobility use, whereas some manufacturers allow continued use down to 70%, and others additionally set limits based on internal resistance growth. This diversity highlights the need for clearer and more harmonized grading criteria across regions and applications. To further enhance classification flexibility and robustness, emerging approaches employ composite scoring systems and soft clustering frameworks that integrate multiple indicators, accommodating differences in chemistry and aging behavior across diverse battery populations [27], [45].

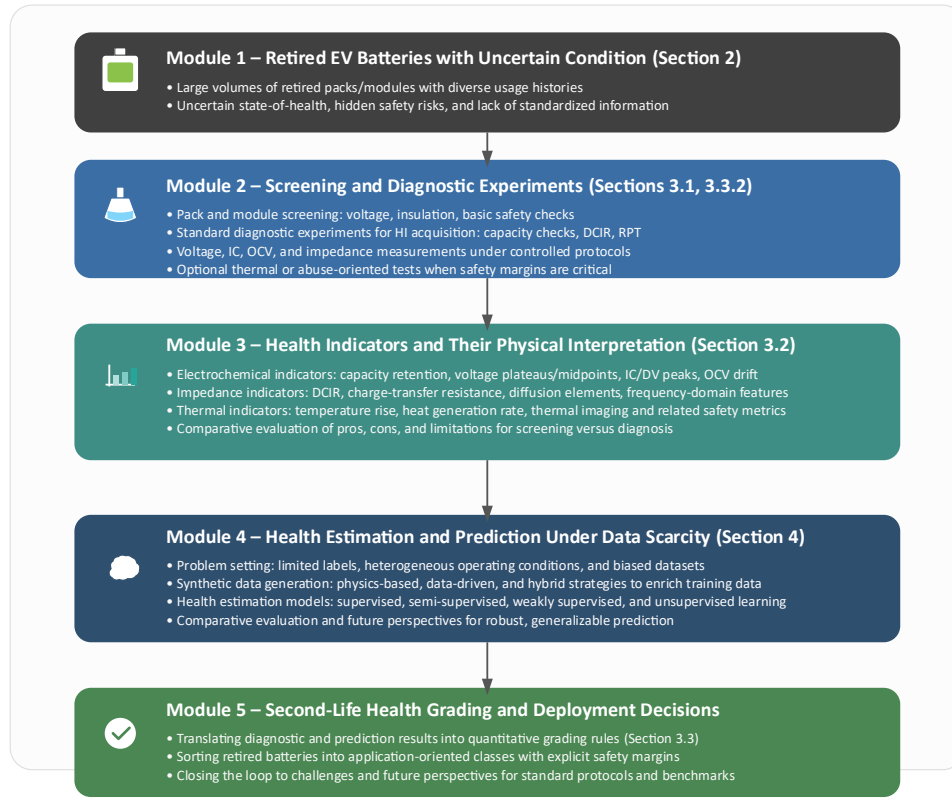

Fig.2 Framework of Retired Lithium-Ion Batteries: from retired EV packs with uncertain condition to screening, indicators, data-scarce prediction, and second-life deployment

Despite notable progress, several persistent challenges remain. Full capacity tests offer the highest reliability in SOH estimation but are too time- and energy-intensive for large-scale use. In contrast, rapid tests, though efficient, are highly sensitive to temperature, state of charge, and procedural variations, resulting in inconsistent measurements across facilities. Additional sources of error stem from variations in current rates, cutoff voltages, ambient conditions, and hardware calibration, all of which undermine cross-laboratory comparability and limit the generalizability of predictive models [46]. These issues highlight the pressing need for harmonized diagnostic protocols, shared reference datasets, and unified reporting standards. The concept of a "battery passport", a digital record encompassing a cell's manufacturing origin, chemistry, and detailed operational history, has been championed by policymakers and industry consortia to improve traceability and facilitate efficient grading for second-life use [47], [48].

In summary, systematic preliminary screening, combining visual inspection, rapid electrical assessment, and targeted advanced electrochemical analysis, forms the backbone of safe and effective R-LIB evaluation. However, practical screening often yields incomplete or inconsistent datasets, especially when integrating information from multiple sources or facilities. This motivates recent research into data generation, completion, and augmentation strategies—topics that will be explored in the following section. The external health indicators identified here provide the foundation for a more detailed discussion of diagnostic metrics and methodologies in the following section.

## 3.2    Health Indicators and Their Physical Interpretation

Health indicators (HIs) constitute the quantitative foundation for assessing the degradation state of R-LIBs. Because the underlying electrochemical aging mechanisms, such as SEI thickening, active-material isolation, impedance growth, and lithium inventory loss, cannot be directly measured at scale, practical evaluation relies on macroscopic signals that serve as interpretable proxies for these microscopic processes. A robust HI framework must therefore balance three critical attributes: physical interpretability, diagnostic sensitivity, and scalability for industrial screening. In this section, we systematically categorize the principal HIs according to their physical origins, outline the mechanisms they reflect, and discuss their applicability and limitations in R-LIB sorting.

### 3.2.1    Electrochemical Performance Indicators

Electrochemical performance indicators describe how an R-LIB responds to applied current and voltage stimuli, thus serving as the most direct macroscopic expressions of its underlying electrochemical state. These indicators primarily capture degradations associated with loss of lithium inventory (LLI), loss of active material (LAM), impedance growth, changes in reaction kinetics, and thermodynamic drift. Compared with purely electrical or thermal indicators, electrochemical performance metrics are more closely tied to fundamental intercalation chemistry, and therefore provide physically interpretable pathways for diagnosing aging mechanisms. This subsection categorizes the major electrochemical performance HIs into capacity-based, voltage-based, derivative-based (IC/DV), OCV-based (equilibrium), and coulombic efficiency indicators.

#### 3.2.1.1    Capacity-based Indicators

Capacity-based indicators form the cornerstone of R-LIB health assessment, as they directly quantify the usable charge a cell can deliver relative to its design specification. The retained capacity—commonly expressed as

$$\text{SOH}_Q = \frac{Q_{\text{measured}}}{Q_{\text{rated}}} \times 100\%.$$

where $Q_{\text{measured}}$ is the dischargeable capacity during diagnostic cycling. Capacity fade arises from two dominant degradation pathways. The first is LLI, primarily driven by SEI thickening, electrolyte reduction, and lithium trapping, which reduce the amount of cyclable lithium and shift electrode stoichiometry. The second pathway is LAM, which results from particle cracking, delamination, or loss of electronic percolation, thereby decreasing the volume of active material participating in intercalation reactions. Seminal degradation studies by Seo et al. [49] and Wang et al. [50] demonstrated that LLI dominates early-life degradation, whereas LAM becomes increasingly pronounced at later stages or under high C-rate cycling.

Although full charge–discharge cycling at low rates (0.2–0.5C) provides the most accurate capacity evaluation, it remains impractical for large-scale R-LIB sorting due to its substantial time and energy cost. This has motivated the adoption of partial-range capacity tests. Despite its universality and direct application relevance, capacity-based SOH alone cannot distinguish between underlying degradation modes, nor can it resolve early or incipient failure mechanisms before substantial loss occurs. For this reason, capacity measurement is often used in conjunction with other electrochemical indicators to achieve a more granular and mechanism-specific health evaluation.

#### 3.2.1.2    Voltage-based Indicators

Voltage-based health indicators can be broadly interpreted as descriptors of the voltage-profile shape. Whether measured during constant-current (CC) charging, constant-power operation, or short relaxation following current interruptions, the voltage trajectory reflects the combined effect of internal resistance, polarization, diffusion kinetics, and phase-transition behavior. Aging distorts the geometry of the voltage curve in predictable ways, such as shifts, slope variations, and local expansions or contractions, making voltage-profile morphology a unifying basis for a wide range of HIs, from localized interval descriptors to global curve-similarity metrics.

Building on this principle, a substantial body of work has explored how localized voltage segments encode degradation signatures. Peng et al. [51] demonstrated that within a targeted voltage interval, aging systematically alters both the rise-rate geometry and localized slope of the CC profile. Their set of interval-based features, including differential-voltage rise time, mean voltage, spectral components, and distance-based deviation metrics, captures the accelerated and sharpened voltage rise that emerges as resistance and polarization increase. Extending the utility of short voltage windows, Li et al. [52] showed that even 300-s CC voltage snippets contain rich statistical traces of degradation: descriptors such as mean, variance, RMS, skewness, and kurtosis of both $V$ and $\Delta V$ were found to vary monotonically with cycling. These studies collectively indicate that fine-grained voltage-shape attributes extracted from short segments can serve as compact yet sensitive SOH indicators.

Beyond localized intervals, global voltage-profile geometry further preserves degradation information. When full CC charging curves are available, Wang et al. [53] demonstrated consistent aging-induced deformation, most noticeably left-shifts and slope steepening, supported by segment-wise statistics and CC–CV transitional behavior. Recognizing that CC data may be unavailable in real-world EV operation, Lai et al. [54] reconstructed an equivalent CC voltage profile from dynamic driving data. Features derived from this reconstructed trajectory, including voltage-rise duration and average voltage-change rate, remained strongly correlated with SOH, illustrating that global voltage-shape characteristics are robust to the means by which the profile is obtained.

Complementing standard CC-based analyses, alternative charging conditions and voltage representations introduce additional diagnostic perspectives. Under fast-charging protocols, the companion study [55] reported accelerated polarization-driven voltage rise, shifts in the effective charging plateau, and pronounced relaxation tails following pulse removal. These dynamic voltage behaviors carry mechanistic significance, reflecting increased impedance, heat accumulation, and potential lithium plating. Moreover, Chen et al. [56] showed that deformation of the voltage–capacity curve—manifested as plateau shifting, stretching or compression, and local slope deviation, can distinguish among different degradation modes such as LLI and LAM. These complementary findings highlight that voltage-profile morphology, expressed either in the time domain or the capacity domain, provides a rich diagnostic signal for multiple aging pathways.

Taken together, these studies demonstrate that voltage-based His, whether derived from local voltage snippets, full or reconstructed CC trajectories, or dynamic fast-charging responses, are unified by the underlying principle of capturing the evolving shape of the voltage profile. This family of indicators provides a physically grounded, model-agnostic means of quantifying aging-induced changes in resistance, lithium inventory, and kinetic limitations, offering one of the most broadly applicable approaches for battery health assessment.

### 3.2.1.3    Derivative-based Indicators

Derivative-based indicators extract degradation signatures by differentiating the voltage–capacity or voltage–time trajectory, thereby amplifying subtle geometric distortions that are difficult to observe directly in raw voltage curves. By operating on the slope and curvature of the electrochemical response, these indicators provide a sensitive view of phenomena such as lithium inventory loss, active material depletion, rising polarization, and the gradual smoothing of phase transitions. Within this unified mathematical perspective, first-order derivatives capture the dominant evolution of plateau slopes and transition-region peaks, whereas higher-order derivatives expose curvature changes that often emerge earlier in the degradation pathway.

First-order derivatives, including incremental capacity (dQ/dV) and differential voltage (dV/dQ), constitute the most widely validated indicators extracted from the V–Q trajectory. Horizontal shifts of IC/DV peaks represent one of the clearest signatures of lithium inventory loss: as SEI growth consumes cyclable lithium, electrode equilibrium potentials drift, and all major IC peaks translate systematically along the voltage axis, a trend consistently observed in mechanistic studies [57], [58], [59]. In contrast, selective reduction of peak amplitude is strongly associated with loss of active material, where particle fracture or electronic isolation reduces accessible transition-region capacity [60], [61]. Recent physics-informed voltage–capacity models likewise reproduce peak translation and plateau distortion as natural consequences of stoichiometric drift [62]. Increasing diffusion resistance and impedance growth further lead to peak broadening, weakening the sharpness of two-phase transitions as cycling progresses [63].

Beyond IC analysis, differential voltage features provide a complementary view of aging. Flattening of DV valleys reflects increasing overpotential and kinetic limitations, particularly at low discharge rates where voltage gradients dominate the response [64]. Because DV features remain identifiable in partial cycles, they have shown strong applicability to EV and operational datasets where full CC charging is unavailable [65]. To mitigate measurement noise, several studies quantify morphological descriptors—such as peak sharpness, curvature, prominence, or inter-peak spacing—which enhance robustness while preserving physical interpretability [50], [66].

Higher-order derivatives offer an even more sensitive probe of curvature evolution, particularly in chemistries with pronounced two-phase behavior. As impedance and diffusion limitations accumulate, curvature around phase boundaries decreases and inflection points are progressively smoothed, revealing early loss of transition-region clarity [67], [68]. Such curvature-driven signatures have been visualized directly in ultrahigh-resolution digital-twin reconstructions, where voltage-curve flattening was shown to precede measurable capacity decay across both LFP and layered-oxide systems [69]. Although second-order derivatives are more susceptible to noise, their strong physical grounding and early-stage sensitivity make them a valuable complement to first-order IC/DV features, particularly in high-fidelity laboratory datasets.

Together, derivative-based indicators form a compact yet physically rich set of health descriptors. First-order derivatives reflect the dominant effects of lithium inventory loss, active material degradation, and kinetic polarization, whereas second-order derivatives provide early sensitivity to curvature changes associated with phase-transition fading. Their combined use enables small geometric distortions in voltage–capacity trajectories to be translated into actionable degradation insights across laboratory, industrial, and field environments.

### 3.2.1.4    Rest and Equilibrium Voltage Indicators

Rest and equilibrium voltage indicators analyze the cell's zero-current voltage response to reveal aging-induced shifts in intrinsic electrode potentials. Unlike voltage-based features obtained under load, these indicators probe the thermodynamic landscape of the electrodes, including their equilibrium potential curves, relaxation trajectories, and dual-path hysteresis behavior, thereby offering direct insight into LLI, LAM, and changes in phase-transition stability. Regardless of whether the inputs are obtained from long-rest OCV measurements, short-term relaxation segments, or paired charge–discharge equilibrium paths, their common foundation lies in observing how the cell approaches or departs from its equilibrium states under I = 0 conditions.

The evolution of the full OCV–SOC curve provides one of the most interpretable equilibrium signatures of degradation in lithium-ion batteries. Progressive horizontal shifts in the OCV trajectory remain the clearest marker of lithium inventory loss, reflecting the altered stoichiometric balance between the two electrodes as SEI growth reduces cyclable lithium [70]. In parallel, slope steepening and plateau compression across characteristic SOC windows have been consistently linked to loss of active material, diminished two-phase coexistence, and weakened phase-transition sharpness in both layered and olivine chemistries [71]. Building on these mechanistic insights, a growing body of work has modeled OCV–SOC evolution directly from low-current equilibrium measurements, demonstrating that analytical expressions fitted across different temperatures and cycling stages capture systematic equilibrium-voltage drifts associated with capacity fade [72]. Other studies have reconstructed pseudo-OCV curves from operational data, either through smooth IC-based regression, online OCV identification, or half-cell OCV composition, to enable real-time tracking of LLI and LAM without requiring full rest conditions [70], [72], [73]. Together, these approaches establish full-equilibrium and quasi-equilibrium OCV curves as reliable and physically grounded indicators whose shape evolution encodes the dominant aging modes over a battery's lifetime.

Short-duration relaxation segments provide an efficient proxy for accessing quasi-equilibrium behavior without requiring full-rest OCV tests. Immediately after current interruption, the voltage decays from its polarized state toward equilibrium, and the shape of this short relaxation window encodes aging-dependent changes in surface polarization and lithium redistribution. Fan et al. showed that even the first 10 s of decay exhibit a monotonic acceleration with cycling, enabling highly accurate SOH estimation directly from the early-time relaxation profile [74]. This trend is reinforced by the 15-s relaxation study of Zhang et al., where time-series features such as curvature and quantile dynamics provide robust degradation signatures without electrochemical model identification [75]. Beyond direct SOH regression, Liu et al. demonstrated that a short relaxation sequence can be mapped to a full V–Q curve through an encoder–decoder framework, enabling access to equilibrium-derived indicators such as IC peaks using only minimal rest time [76]. Together these results confirm that relaxation-based indicators offer a compact yet information-rich representation of the equilibrium trajectory, well suited for real-world applications where natural rest periods are abundant.

Charge–discharge hysteresis provides an additional view of equilibrium behavior by contrasting two quasi-equilibrium voltage paths traced under identical currents but opposite lithiation directions. The evolution of the hysteresis loop, whether observed as narrowing due to stoichiometric shifts associated with LLI or as widening resulting from mechanical strain or sluggish phase transitions, has been repeatedly confirmed in mechanistic studies examining entropy-based hysteresis signatures and their sensitivity to aging and cycling direction [77]. Building on these physical insights, detailed investigations of SOC-, temperature-, and path-dependent hysteresis behavior in both LFP and ternary cathode systems have demonstrated that the loop shape, asymmetry, and charge–discharge voltage separation encode rich information about electrode thermodynamics, motivating asymmetric hysteresis operators capable of accurately capturing these dependencies[78]. At the same time, recent efforts in data-efficient OCV-envelope reconstruction have shown that only sparse measurements of the hysteresis window are required to recover the underlying equilibrium trajectories, enabling accurate aging-state adaptation and cross-chemistry transfer of hysteresis models without intensive parameterization [79]. Because hysteresis captures directional equilibrium effects that remain invisible in single OCV–SOC curves or short relaxation segments, it offers complementary diagnostic value for interpreting electrode-level degradation and remains particularly informative in materials exhibiting pronounced two-phase transitions such as LFP.

Together, rest-derived relaxation features, full OCV–SOC trajectory evolution, and dual-path hysteresis shifts form a coherent family of equilibrium-rooted indicators that probe electrode thermodynamics more directly than voltage-based or impedance-based metrics. Their shared advantage lies in strong interpretability, each reflects specific degradation mechanisms at the electrode level, while their practical differences stem from testing burden: full OCV provides the richest information but is difficult to obtain, relaxation segments offer excellent deployability, and hysteresis reveals electrode asymmetry inaccessible to other modalities.

### 3.2.2    Impedance Indicators

Impedance and transport indicators provide a deeper and more mechanistically resolved view of degradation in retired lithium-ion batteries than capacity or voltage-based metrics. Whereas electrochemical performance indicators reflect the macroscopic outcome of aging, impedance descriptors probe the underlying resistive, capacitive, and diffusive processes governing charge transfer and lithium-ion mobility. Because these physicochemical pathways are directly altered by SEI growth, electrolyte degradation, particle cracking, and transition-metal dissolution, impedance- and transport-based health indicators have become essential tools for diagnosing the origin and severity of degradation. This subsection synthesizes the major categories of impedance and transport indicators, direct-current resistance, frequency-domain impedance spectroscopy, relaxation-time analysis, and diffusion-related metrics, and discusses their mathematical formulations, physical interpretations, and relevance to R-LIB screening.

### 3.2.2.1    Direct-Current Internal Resistance (DCIR)

Direct-current internal resistance (DCIR) is among the most practical and widely adopted impedance-based health indicators owing to its simplicity, fast measurement, and low instrumentation requirements. As a lumped representation of ohmic resistance, interfacial charge-transfer kinetics, and early-stage diffusion polarization, DCIR reliably increases as the cell ages, making it an accessible surrogate for cumulative impedance buildup. For automotive-grade lithium-ion batteries, DCIR is routinely monitored alongside capacity fade to track degradation across diverse chemistries, usage patterns, and environmental conditions.

Recent experimental studies have reinforced the strong connection between cycling-induced degradation and DCIR evolution. Investigations on prismatic NMC/graphite cells operated under fast-charging regimes have shown clear monotonic increases in DCIR, reflecting the combined influence of rate-dependent interfacial degradation and cathode structural changes that emerge under aggressive charging conditions [80]. Temperature has also been identified as a dominant external factor shaping DCIR progression: controlled cycling experiments on LFP and NMC cells demonstrate that elevated operating temperatures accelerate resistance rise, whereas active cooling can substantially suppress DCIR drift and nearly double the attainable cycle life [81]. These results emphasize that DCIR is sensitive not only to intrinsic electrochemical aging but also to thermal environment and discharge-rate conditions.

Beyond individual aging studies, DCIR is now widely incorporated into large-scale experimental designs as a core diagnostic variable for modeling degradation behavior. Multi-stage aging datasets involving hundreds of cells under systematically varied cycling and storage conditions consistently include DCIR measurements, underscoring its value as a generalizable indicator across chemistries and test protocols [82]. In the context of retired or second-life batteries, DCIR has further gained traction as a rapid screening metric to assess state-of-health variations across large populations. Statistical analyses of LiFePO$_4$ cells harvested from hybrid-electric vehicle packs have demonstrated strong correlations between DCIR and retained capacity, enabling practical acceptance criteria for module-level sorting and highlighting the role of resistance-based metrics in repurposing workflows [83].

Despite these advantages, DCIR provides only a coarse decomposition of resistive processes: increases may arise from SEI growth, loss of active material, charge-transfer degradation, electrolyte aging, or cathode microstructural changes, all of which contribute similarly to the aggregate resistance trend. Its strong dependence on temperature and state-of-charge further necessitates standardized measurement conditions or compensation models to ensure cross-cell comparability. Nevertheless, the ease of measurement and module-level accessibility make DCIR a cornerstone indicator in large-scale diagnostics, production quality control, and second-life battery triage, where rapid decision-making is often more critical than mechanistic specificity.

### 3.2.2.2    Frequency-Domain Impedance Spectroscopy (EIS)

Electrochemical impedance spectroscopy (EIS) offers a frequency-resolved view of battery degradation that is far more interpretable than lumped DC resistance. By separating ohmic, charge-transfer, double-layer, and diffusion contributions, EIS provides a structured way to attribute impedance growth to specific physical mechanisms rather than aggregating them into a single scalar value. This decomposition has made EIS one of the most informative laboratory-grade tools for understanding how different degradation modes unfold over the battery lifetime.

A substantial body of research has characterized the evolution of mid-frequency charge-transfer features, which serve as sensitive markers of interfacial degradation. Increases in the semicircle diameter or extracted $R_{ct}$ parameters have been linked to SEI thickening, surface reconstruction, and transition-metal dissolution in both half-cell and full-cell configurations, as demonstrated in electrolyte-engineering studies and mechanistic aging analyses [84], [85]. Complementary work has examined frequency-domain signatures associated with diffusion and porous-electrode transport. Distortions in the low-frequency tail, including changes in diffusion impedance and relaxation-time distributions, have been reported during diffusion-limited intercalation, pore clogging, and loss of active-material connectivity [84], [86].These findings establish EIS as a powerful tool for distinguishing between interfacial and transport-dominated degradation pathways—capabilities that cannot be achieved with DCIR alone.

EIS features also vary systematically with chemistry, operating conditions, and environmental stress. Studies incorporating temperature and state-of-charge variation have shown that both SEI resistance and charge-transfer kinetics exhibit distinct dependencies on thermal environment and SOC windows, with measurable implications for SOH estimation [87]. Comparative analyses across datasets and battery types have further revealed that impedance spectra carry chemistry-specific fingerprints, enabling cross-cell discrimination and supporting impedance-driven state estimation frameworks [88].

In parallel, recent work has leveraged EIS as a feature-rich input for machine-learning-based diagnostics. High- and mid-frequency geometric features, equivalent-circuit parameters, ZPG model poles, and convolutionally extracted latent features have all been used to improve SOH estimation, achieving errors below 1–3% across a range of temperatures and cell types [89], [90], [91], [92]. These approaches highlight EIS's potential as an interpretable and information-dense input for data-driven prognostics, provided that computational cost and measurement variability are appropriately managed.

Despite these advantages, practical deployment of EIS remains limited by its instrumentation requirements, long stabilization times, and sensitivity to measurement protocol. Equivalent-circuit fitting remains inherently non-unique, introducing operator-dependent variability and limiting cross-study comparability. As a result, EIS is less suitable than DCIR for high-throughput retired-battery sorting, though it remains indispensable for mechanistic aging studies, calibration of physics-based models, and the development of interpretable machine-learning estimators.

### 3.2.2.3 Distribution of Relaxation Times (DRT)

The distribution of relaxation times (DRT) enhances the interpretability of impedance spectra by decomposing the measured response into a continuum of characteristic timescales. Unlike conventional equivalent-circuit fitting, which often suffers from parameter coupling and non-uniqueness, DRT can cleanly separate overlapping electrochemical processes such as charge-transfer kinetics, double-layer charging, solid-state diffusion, and particle–contact effects. This makes DRT particularly powerful for diagnosing aging mechanisms that distort the semicircle structure of Nyquist plots but leave recognizable relaxation signatures in the time domain. Recent studies have shown that DRT can reveal the evolution of interfacial and transport processes during cycling, clarifying how SEI growth, loss of active material, or electrode architecture changes contribute to impedance rise in LIBs [93], [94].

DRT is especially useful for analyzing heterogeneous or aged cells, where spatial inhomogeneities, originating from particle-size variations, uneven SEI morphology, or non-uniform current-collector contacts, broaden or shift the relaxation spectrum beyond what classical EIS can resolve. Comparative DRT studies on commercial NCM, LFP, and NCA cells demonstrated that each chemistry exhibits distinct relaxation peaks associated with different kinetic regimes, enabling a more reliable discrimination of degradation pathways across materials [94]. Similarly, real-world EV aging investigations reported that DRT can decouple cathode charge-transfer impedance from anode contact resistance, providing a mechanistic view of how mileage-dependent transitions occur from lithium-inventory loss toward active-material degradation [95].

Beyond mechanism identification, DRT has become an important tool for health-indicator construction and SOH estimation. Machine-learning approaches that combine DRT peak features with regression models, such as multilayer perceptrons or Gaussian-process regressors, have achieved significantly improved robustness compared with single-frequency EIS features, especially under varying temperature or SOC conditions [96]. Recent developments also extend DRT analysis to all-solid-state batteries and low-temperature impedance characterization, where separating interfacial processes is critical; these studies highlight the ability of DRT to identify electrode/electrolyte contact, particle–particle transport, passive-film growth, and other interfacial phenomena across chemistries and thermal environments [93], [97].

Despite these advantages, DRT requires high-resolution impedance measurements and appropriate regularization, and peak assignment may still require expert interpretation. Nevertheless, by revealing subtle early-stage changes invisible in traditional EIS, and by providing a systematic process-resolved view of impedance evolution, DRT has become one of the most informative impedance-based indicators for retired-battery assessment, degradation mechanism identification, and data-driven health estimation.

### 3.2.3    Thermal Indicators

Thermal indicators characterize how a cell generates, accumulates, and redistributes heat during operation or rest, thereby providing an integrative reflection of internal resistance buildup, parasitic reactions, and structural or interfacial degradation. Many degradation modes, such as impedance rise, electrolyte depletion, cathode micro-cracking, transition-metal dissolution, gas evolution, and micro-short formation, manifest earlier or more sensitively in thermal behavior than in purely electrochemical signatures. Consequently, temperature-based indicators play a critical role in assessing the safety and reliability of R-LIBs, where cells with similar capacity may exhibit very different thermal fragility due to heterogeneous aging histories [98], [99]. These indicators therefore complement electrochemical HIs by revealing latent risks and guiding acceptance decisions for second-life use.

### 3.2.3.1    Temperature Rise and Heat-Generation Indicators

Temperature-rise indicators quantify how rapidly and intensely a cell warms under controlled cycling or pulse excitation, offering a direct measurement of heat generation. The total heat-generation rate arises from irreversible Joule heating and reversible entropic contributions, both of which evolve with aging. SEI thickening, loss of active surface area, electrolyte depletion, and rising charge-transfer resistance all increase the I²R component, whereas structural changes in layered oxide cathodes modify entropic heat through shifts in phase equilibria and $\partial U/\partial T$ behavior[22], [100].

Experimental studies consistently show that aged NMC and LFP cells exhibit faster and larger temperature rise even under moderate cycling. Elevated peak ΔT correlates strongly with internal-resistance growth, lithium plating severity, and the onset of parasitic reactions, indicating that temperature rise is a sensitive marker of early degradation [101], [102]. High-temperature aging studies further reveal that thermal behavior deteriorates systematically with cycling: oxidized cathode structures, thicker or decomposed SEI, and gas formation collectively reduce thermal stability, lower self-heating onset temperature, and accelerate heat accumulation [22].

Because temperature-rise tests can be conducted using short pulses and simple surface-temperature sensors, without full cycling, they offer a rapid, low-cost screening indicator for large retired-battery populations. However, thermal indicators are highly sensitive to external conditions such as cooling environment, cell orientation, and fixture thermal paths; therefore, robust normalization or standardized testing environments are required to ensure comparability across facilities. Even so, temperature-rise indicators remain one of the most practical and safety-relevant tools for detecting thermal fragility and identifying cells unsuitable for second-life deployment.

### 3.2.3.2 Temperature Uniformity and Surface Hot-Spot Indicators

Spatial thermal behavior provides another layer of diagnostic insight beyond bulk temperature rise, particularly for retired batteries where degradation is often highly localized. Infrared (IR) thermography and distributed sensor technologies consistently reveal that non-uniform aging leads to characteristic hot-spot patterns on the cell surface. These local temperature anomalies typically arise from elevated impedance regions, micro-shorts, gas-pocket formation in pouch cells, current-collector corrosion, or tab/weld defects—conditions that may not be captured by global electrochemical indicators alone.

A common metric for quantifying spatial thermal non-uniformity is the temperature spread, defined as the difference between the hottest and coldest regions on the cell surface. Larger spreads indicate highly asymmetric heat generation and are strongly associated with structural inhomogeneity and non-uniform degradation. Recent work using IR thermography has demonstrated that temperature-curve similarity can also serve as a screening metric, enabling single-cycle consistency assessment of retired modules by comparing spatial thermal signatures across cells [103]. Experimental studies further confirm that IR imaging reliably identifies abnormal heat-generating regions that correlate with electrical behavior and can quantitatively approximate heat-generation rates [104].

Beyond surface imaging, emerging distributed sensing technologies provide much finer spatial resolution. Fiber-optic distributed temperature sensing (DTS/DFOS) has been used to track in-plane gradients and the movement of hottest regions during operation, revealing that internal temperature differences can exceed surface-measured values by more than 300% in degraded pouch cells [105]. Such methods offer unique visibility into subsurface thermal heterogeneity, an important advantage for stacked electrodes where surface emissivity and heat-spreading effects mask internal hot spots [106].

In the context of second-life screening, sustained hot-spot behavior is especially concerning because it predicts both accelerated local aging and an elevated risk of thermal propagation. Studies on aged LFP cells, for example, show that internal hot spots can trigger exothermic reactions at temperatures as low as 55 °C, substantially reducing the onset threshold for thermal runaway [107]. Thermal fragility of this kind may disqualify a battery from repurposing even when its electrical performance metrics remain within acceptable bounds. Thus, temperature uniformity indicators, whether obtained through IR imaging or advanced distributed sensing, play an essential role in ensuring safe and reliable reuse of heterogeneous R-LIB populations.

### 3.2.3.3 Runaway-Related Thermal Indicators

Safety-critical thermal indicators focus on early-stage precursors of thermal runaway, capturing chemical instabilities that often emerge well before catastrophic failure. These indicators are particularly important for retired batteries, in which aging-induced heterogeneity, lithium plating, and internal defects substantially narrow the thermal safety margin even if capacity or impedance metrics appear acceptable.

A central precursor is self-heating at rest, which reflects parasitic reactions, micro-short pathways, or unstable SEI decomposition. Healthy cells show negligible temperature drift, while aged or plated cells can exhibit measurable positive temperature rise during rest or mild heating. Several thermal-runaway experiments report that self-heating in NMC, LFP, and LCO cells intensifies during the transition from heat-conduction to gas-generation-dominated phases, serving as one of the earliest measurable warning signals before voltage collapse or sharp temperature acceleration [65], [108].

A second widely used indicator is the exothermic onset temperature, typically extracted from ARC or DSC measurements. Aging, overcharge, lithium plating, and electrolyte decomposition all reduce this onset temperature, shrinking the buffer between normal operation and thermal runaway. Multiple studies show that aged or overcharged cells can enter exothermic stages 20–40 °C earlier than fresh cells, with plated lithium in particular causing dramatically earlier eruption behaviors and intensified gas production [65], [107], [109].

A third precursor is low-temperature gas generation, often detectable between 40–70 °C. Gas evolution at such low temperatures indicates unstable SEI layers, electrolyte oxidation, or early-stage internal shorting. Overcharge-induced TR studies demonstrate that detectable gas peaks and surface swelling occur in early stages, aligning closely with the transition into higher-hazard TR phases [110], [111]. Gas-based or multi-sensor early-warning strategies have been shown to provide tens of minutes of lead time before full runaway, underscoring the diagnostic value of gas evolution combined with temperature indicators [112].

Recent work further highlights that combined thermal–electrochemical signatures enhance early-stage detection. The evolution of characteristic impedance peaks during controlled heating provides complementary precursors to temperature-based cues, enabling two-level or multi-level early warning frameworks that outperform temperature-only approaches [108]. Similarly, thermal-runaway analyses under different aging pathways show that lithium plating, SEI rupture, and oxygen release from layered oxides leave distinct thermal fingerprints, emphasizing the need to interpret thermal precursors in the context of degradation mechanisms [113], [114].

For second-life qualification, these precursors are especially valuable because they target failure mechanisms rather than performance decline. Even cells with seemingly acceptable capacity or DCIR may show early gas generation, reduced onset temperature, or abnormal resting heat—conditions that indicate unacceptable safety risks for repurposing. As such, safety-critical thermal indicators serve as decisive exclusion criteria for R-LIB screening and represent an essential layer of protection against catastrophic events in second-life deployments.

3.2.4    Comparative Evaluation and Limitations

The health indicators reviewed across Sections 3.2.1–3.2.3 capture different but complementary manifestations of degradation and collectively form the foundation for screening and diagnosing retired lithium-ion batteries. Table 1

## Table 1 — Comparative summary of major indicator categories

| HI Category | Representative Features | Physical Meaning | Measurement Requirements | Strengths | Limitations |
|---|---|---|---|---|---|
| Electrochemical indicators (3.2.1) | Capacity, voltage profile shape, IC/DV peaks, OCV behavior | Stoichiometry shifts (LLI), active-material loss, phase-transition changes, polarization | Controlled cycling (CC/CV), relaxation steps for OCV, high-resolution voltage/capacity data | High interpretability; widely validated; strong correlation with capacity/SOH | Testing time is long; sensitive to protocol and temperature; noise affects IC/DV |
| Impedance-based indicators (3.2.2) | DCIR, high-/mid-/low-frequency EIS, DRT components | Ohmic resistance, charge-transfer kinetics, SEI/interfacial aging, ion transport limitations | Pulse or frequency sweep tests; SOC & temperature control; specialized instrumentation | Mechanistic resolution; early-stage degradation detection; fast when automated | Requires dedicated hardware; SOC/temperature dependence; complex interpretation |
| Thermal & safety indicators (3.2.3) | Temperature spread (ΔT), hot-spot intensity, self-heating rate, gas evolution, DSC/ARC onset temperature | Heat generation from reaction kinetics, micro-shorts, SEI instability, runaway precursors | IR imaging, sensor arrays, rest-period monitoring, calorimetry, controlled heating | Essential for safety; detects latent instabilities; identifies structural defects | May require long rest/heating times; emissivity errors; internal temperatures not directly observable |

synthesizes these indicators by comparing their physical meaning, representative features, measurement requirements, diagnostic resolution, and typical limitations. This consolidated perspective clarifies how each indicator class contributes to core assessment tasks such as capacity estimation, power capability evaluation, fault identification, and safety assurance.

Electrochemical performance indicators, particularly capacity-, voltage-, and derivative-based metrics, provide high interpretability and direct links to material stoichiometry and reaction pathways. They remain the gold standard for quantifying usable capacity, although their dependence on controlled charge–discharge cycling makes them time-intensive for large-scale screening. Impedance-based indicators offer finer mechanistic resolution by distinguishing ohmic, interfacial, and diffusion processes, but require specialized instrumentation and tight control of temperature and state of charge. Thermal indicators and self-discharge metrics reveal latent instability and parasitic reactions that may not appear during short cycling tests, but they typically require long rest periods or calorimetry-based setups.

Overall, no single indicator provides a complete representation of battery health. Effective R-LIB evaluation requires balancing diagnostic accuracy, scalability, and practical constraints. Multi-indicator fusion，combining the mechanistic insight of electrochemical and impedance metrics with the safety sensitivity of mechanical and thermal indicators, offers the most robust foundation for large-volume repurposing, ensuring both performance reliability and safety compliance.

## 3.3 Diagnostic Techniques and Practical Health Grading

### 3.3.1 Purpose and Scope of Diagnostic Techniques

Diagnostic techniques provide the experimental foundation for converting conceptual health indicators into measurable, comparable, and decision-ready quantities. While Section 3.2 summarized the electrochemical, impedance, thermal, and mechanical indicators that describe battery aging, these indicators become operationally meaningful only when acquired through standardized diagnostic procedures. For R-LIBs, where cell populations originate from diverse usage histories and require rapid but reliable evaluation, diagnostic testing must simultaneously achieve accuracy, reproducibility, and high throughput.

To meet these needs, a range of experimental techniques has been developed, from full charge–discharge profiling and pulse-based resistance measurements to impedance spectroscopy, thermal characterization, and emerging rapid-test protocols. Each method probes a different subset of aging mechanisms, yielding complementary data streams that together form the basis for practical health grading. These measurements are subsequently processed, normalized, and mapped into quantitative decision rules that support large-scale sorting, application routing, and safety assurance in second-life deployment.

This section reviews the principal diagnostic methods used to obtain health indicators in practice and outlines how their outputs are transformed into consistent grading frameworks. By establishing this experimental-to-decision workflow, Section 3.3 bridges the physics-based interpretation of aging (Section 3.2) with the data-driven modeling approaches discussed in Section 4.

### 3.3.2 Standard Diagnostic Experiments for HI Acquisition

#### 3.3.2.1 Full and Partial Charge–Discharge Tests

Full charge–discharge profiling is the most widely used and physically transparent diagnostic method for evaluating retired lithium-ion batteries. By cycling cells under controlled constant-current conditions, reference voltage–capacity curves can be obtained to quantify retained capacity, Coulombic efficiency, plateau characteristics, overpotential evolution, and derivative-based indicators such as IC and DV curves. These signatures collectively reflect lithium inventory loss, active-material degradation, and increasing polarization during aging, making full-cycle profiling a benchmark approach for establishing ground-truth SOH in both laboratory studies and second-life assessments [115], [116].

Nevertheless, full cycling is time-consuming and not always feasible for large inventories of retired cells. To improve screening throughput, recent studies have explored partial-range charge–discharge protocols, which extract diagnostic features from only the most informative voltage windows. For LiFePO$_4$ (LFP) cells, IC peaks and plateau slopes obtained from mid-voltage or high-rate charging segments exhibit strong correlation with the true usable capacity, allowing for rapid feature acquisition at a fraction of the time required by a full cycle [117], [118]. Partial charging segments also enable extraction of multiple features, such as voltage rise rate, internal resistance proxies, and localized curve deformation, providing a cost-effective basis for fast sorting schemes [42].

Derivative-based indicators extracted from these profiles further enhance diagnostic sensitivity. Comparative analyses have shown that the quality of IC/DV curves, their peak alignment, and the robustness of numerical differentiation strongly influence the reliability of SOH estimation [119]. Coupling IC/DV signatures with full or partial cycling data helps reveal subtle aging phenomena such as phase-transition smoothing and shifting equilibrium potentials, complementing capacity-based measurements [120].

Together, full and partial charge–discharge profiling form the foundation of practical diagnostic testing for retired batteries, offering a balance between physical interpretability, test accuracy, and operational efficiency.

### 3.3.2.2    Hybrid Pulse Power Characterization and DCIR Tests

Pulse-based diagnostic techniques provide rapid and sensitive assessment of internal resistance and power capability, making them indispensable for large-scale screening of retired lithium-ion batteries. In HPPC, a sequence of controlled charge–discharge pulses is applied while monitoring the instantaneous voltage drop. The resulting DCIR reflects combined contributions from ohmic, charge-transfer, and diffusion overpotentials, and it evolves systematically with SOC and SOH. Because resistance growth often precedes noticeable capacity loss, pulse-derived resistance metrics serve as reliable early indicators of degradation, particularly in second-life applications where rapid sorting is required [121], [122].

Recent studies demonstrate that more advanced pulse designs can significantly enrich diagnostic information. For instance, injecting high-frequency or multi-level pulses enables the extraction of pulse harmonics, which carry signatures related to kinetic parameters, electrode phase transitions, and nonlinear open-circuit voltage contributions. Machine-learning models trained on pulse harmonics can reconstruct IC extrema and estimate SOH with sub-percent errors, achieving both impedance-like and IC-like diagnostics within a few minutes [121], [123]. These findings indicate that pulse harmonics can simultaneously capture the effects of ohmic losses, charge-transfer resistance, and diffusion limitations, positioning pulsing as a promising bridge between IC/DV analysis and impedance spectroscopy.

Variations of the HPPC protocol have also been explored to optimize identification of equivalent-circuit model parameters. Systematic analysis of pulse height, duration, and relaxation time reveals that pulse configuration has a measurable impact on parameter identifiability under dynamic and quasi-static conditions [123]. Such optimization is particularly relevant for screening retired batteries with heterogeneous aging histories, where fast but physically consistent parameter extraction is required.

Pulse-based diagnostics are further applied in extreme usage conditions, such as ultrahigh-rate discharge or high-C-rate operation, where pulse-derived voltage responses reveal degradation mechanisms linked to loss of active material, lithium inventory loss, or plating-induced impedance rise [49], [124]. These studies highlight the versatility of pulse-based methods in capturing degradation modes that conventional full-cycle tests may overlook.

Overall, pulse-based resistance and power tests provide a practical balance between diagnostic depth and operational efficiency. Their short test duration, compatibility with automated screening equipment, and sensitivity to early degradation make them particularly suitable for second-life battery sorting, where thousands of cells must be evaluated rapidly yet reliably.

### 3.3.2.3    EIS

EIS has become a central diagnostic technique for evaluating degradation in R-LIBs. Its appeal lies in its rapid acquisition, non-destructive nature, and sensitivity to kinetic, interfacial, and transport processes that evolve throughout aging. In practical experimental settings, particularly during large-scale sorting and repurposing, EIS-derived health indicators provide a unique balance between physical interpretability and operational efficiency.

A dominant experimental approach involves fitting the measured Nyquist and Bode spectra to equivalent circuit models (ECMs). The resulting parameters, such as SEI resistance, charge-transfer resistance, or diffusion-related impedances, show strong monotonic trends with capacity loss and interfacial aging. Recent work has demonstrated that augmenting traditional Randles circuits with additional capacitive branches can improve the identifiability of SEI-related processes and provide more robust inputs for data-driven SOH estimators under varying temperatures [125]. More mechanistically grounded ECMs, such as transmission-line circuits and frequency-dispersive Warburg components, have been used to capture porous-electrode transport more faithfully, enabling higher-fidelity interpretation in materials with complex diffusion pathways [126]. These models, however, require careful parameter initialization and may suffer from non-uniqueness in high-noise or highly aged conditions, limiting their robustness for industrial-scale retired-battery testing.

To mitigate model-dependence, machine-learning methods that operate directly on raw spectra have gained momentum. By training on large experimentally collected EIS datasets covering wide ranges of SOC, temperature, and aging states, Gaussian process and neural network models have demonstrated the ability to automatically identify degradation-relevant spectral regions without prior circuit assumptions [127]. Such approaches significantly reduce the burden of model fitting and are more tolerant of spectrum distortions commonly observed in retired battery samples. Their practical limitation, however, lies in the need for comprehensive training datasets that capture real-world variability across chemistries, formats, and historical usage patterns.

Complementing ECM and ML approaches, the distribution of relaxation times (DRT) has emerged as a powerful tool for decomposing overlapping physicochemical processes. When applied in experimental comparative studies across different commercial chemistries, DRT has effectively revealed distinct time-constant signatures associated with SEI growth, charge-transfer kinetics, and solid-state diffusion [94]. Such decomposition provides richer interpretability than Nyquist plots and facilitates cross-cell comparison, supporting practical decisions such as chemistry-specific sorting thresholds and identification of anomalous cells within heterogeneous retired packs.

From the perspective of large-scale application, impedance-derived indicators have been integrated into rapid sorting and regrouping workflows. Short-time measurements combining EIS with dynamic voltage features have enabled cell-consistency assessments within minutes, offering a viable pathway for automated industrial screening lines [36]. Similarly, simplified ECM parameters extracted at a limited number of frequencies have been employed for rapid SOH prediction in decommissioned power batteries, achieving estimation accuracy below 2% while dramatically reducing testing time [128]. These studies illustrate that, when appropriately simplified or combined with complementary dynamic signals, impedance-based HIs are fully compatible with high-throughput R-LIB sorting environments.

Overall, impedance-based indicators offer a technically rigorous yet operationally efficient pathway for R-LIB assessment. Their strengths include high sensitivity to early degradation, compatibility with diverse cell formats, and ability to extract mechanistic signatures unobtainable from voltage or capacity alone. Nonetheless, practical deployment faces persistent challenges: ensuring consistent SOC/temperature during testing, mitigating parameter non-uniqueness in ECM fitting, and managing data requirements for ML-based methods. Future efforts are expected to integrate real-time EIS acquisition with adaptive modeling frameworks and standardized test protocols to enable reliable, scalable health assessment for second-life battery applications.

### 3.3.2.4 Thermal and Safety Diagnostics

Thermal characterization provides an essential experimental window into degradation and safety phenomena in retired lithium-ion batteries, since many aging mechanisms, including resistance growth, SEI thickening, lithium plating, and gas generation, directly alter heat generation and temperature evolution. Surface-temperature measurements during controlled charge–discharge cycling remain one of the most practical methods. By attaching thermocouples or NTC sensors to the cell surface, researchers track temperature rise rates, cooling behavior, and transient inflection points that reflect internal resistance changes and electrochemical reaction shifts. Experimental studies have shown that the discharge-temperature trajectory and its fitted parameters evolve consistently with aging [99], while differential temperature curves during charging generate repeatable thermal signatures tied to degradation progression [129].

Thermal response becomes even more informative under accelerated or abusive conditions. Overcharge tests on LiFePO$_4$ and graphite cells reveal distinct thermal stages involving gas evolution, separator degradation, and reaction acceleration, each identifiable through characteristic temperature rise patterns and onset temperatures of abnormal heating [110], [130]. Such experiments provide sensitive early indicators of compromised cells that may not be detectable from voltage measurements alone. Complementary thermo-mechanical sensing approaches, such as measuring expansion force during overcharge, have shown that abnormal mechanical deformation can precede surface-temperature anomalies by hundreds of seconds, offering an even earlier warning of runaway-related processes [131].

More advanced diagnostic setups employ spatially resolved thermal sensing. Fiber Bragg Grating (FBG) sensors and distributed fiber-optic systems enable mapping of localized or internal temperature gradients with higher temporal and spatial resolution than conventional surface sensors. Experimental comparisons confirm that FBG sensing captures sharper thermal gradients and earlier abnormal rises during overcharge [132], while embedded distributed fiber-optic measurements reveal axial thermal inhomogeneities within cylindrical cells that reflect internal structural aging [133]. These high-resolution techniques, though more complex, provide valuable insight into degradation-induced thermal asymmetry.

Calorimetric and hybrid thermal-behavior experiments further support thermal diagnostics by quantifying heat-generation rates under realistic operating currents. Combined electrical–thermal experimental platforms have demonstrated strong agreement between measured and inferred heat generation, validating the use of thermal signatures as physically grounded indicators of aging and abnormal reaction pathways [134].

Collectively, these thermal diagnostic experiments, from simple surface thermometry to advanced fiber-optic sensing and controlled overcharge tests, form a robust suite of tools for extracting degradation-relevant thermal indicators. Their practicality, sensitivity, and close linkage to safety-critical mechanisms make them indispensable for large-scale screening and grading of retired batteries.

### 3.3.3 Translating Diagnostic Results into Quantitative Grading Criteria

Practical health grading transforms raw diagnostic measurements into actionable decisions that determine how retired lithium-ion batteries are reused, routed, or rejected. The process begins with converting electrical, thermal, and structural test outputs into standardized health metrics. Routine measurements, such as retained capacity from partial or full discharge tests, DCIR from pulse characterization, impedance ratios from rapid perturbation methods, voltage midpoint drift or plateau-slope deformation from voltage curves, and thermal-rise coefficients observed during load excursions, are each mapped to specific degradation modes. This mapping enables different classes of health indicators to be interpreted in a consistent, operational form suitable for large-scale screening.

Building on these metrics, industrial grading schemes generally adopt either fixed thresholds or continuous scoring. Threshold-based methods classify cells according to minimum requirements on capacity, internal resistance, or thermal stability; continuous scoring constructs composite indices that weight energy capability, power capability, and safety margins according to application needs. For example, stationary storage typically prioritizes capacity retention, whereas mobility applications favor power-oriented metrics such as DCIR or polarization resistance. Recent studies further show the benefit of integrating multiple health features—capacity-related indicators, resistive characteristics, and polarization behavior, into low-dimensional health vectors that improve the consistency of sorting decisions [135]. Similarly, resistive remaining capacity and cross-validated SOH estimation have been used to reduce the sorting time of cells with unknown or variable SOC conditions [33]. Across these approaches, grading is not merely an SOH estimation exercise; it is a routing decision that determines which second-life domain a cell is suitable for.

Scaling these decisions to industrial throughput relies increasingly on automated sorting lines. Modern facilities deploy multi-station workflows in which cells undergo a short voltage scan, rapid DCIR or pulse response measurement, thermal sensing, and visual inspection. These systems normalize incoming measurements in real time and compare them against tolerance envelopes calibrated from validated benchmarks. High-speed implementations, such as multi-frequency electrical excitation platforms capable of diagnosing cells within one second, demonstrate how rapid perturbation techniques can enable gigafactory-scale or warehouse-scale evaluation without lengthy relaxation or cycling requirements [136]. Automated optical inspection further contributes by identifying surface defects, swelling, leakage traces, or weld/tab abnormalities that reflect safety-critical mechanical degradation.

Finally, effective grading requires calibration and harmonization across laboratories and facilities. Differences in test protocols, temperature management, sampling resolution, and equipment sensitivity can produce noticeable variance in measured health indicators. Cross-facility discrepancies have been highlighted both in safety-oriented diagnosis [137] and in broader battery development roadmaps emphasizing the need for interoperable diagnostic standards and shared data infrastructure [138]. Emerging industrial practices now integrate statistical calibration, data-driven adjustment of decision thresholds, and cross-validated benchmarks to reduce variability and ensure reproducible routing outcomes. These considerations naturally connect to the next section, where diagnostic data completion, harmonization, and synthetic augmentation are discussed as strategies for further improving grading transferability.

## 4 Health Estimation and Prediction Under Data Scarcity

The health estimation of R-LIBs presents fundamentally different challenges from those encountered in new or in-use batteries. As summarized in Section 3, individual HIs, capacity retention, DCIR, impedance features, voltage or IC curve characteristics, thermal behavior, and others, offer valuable but incomplete perspectives on degradation. However, retired batteries entering second-life screening typically exhibit high data sparsity, missing or noisy labels, unknown historical usage, and large cell-to-cell heterogeneity arising from diverse chemistries, aging trajectories, and operational stresses. These constraints not only weaken the diagnostic reliability of any single HI, but also limit the applicability of conventional laboratory protocols, which rely on controlled cycling, long stabilization periods, or full electrochemical characterization. As a result, practical health prediction at the retirement stage requires methods capable of learning from incomplete, irregular, and minimally measured data.

To address these limitations, recent studies have increasingly explored synthetic data generation as a means to augment real screening measurements and compensate for missing degradation trajectories. Synthetic data can be produced through physics-based models, data-driven generative methods, or hybrid approaches that embed physical constraints into learned latent spaces. These generated samples help enrich the distribution of degradation patterns, improve model robustness, and alleviate the dependence on long-term cycling data, an essential requirement in industrial retired-battery sorting.

With both real and generated data available, a broad spectrum of health estimation and prediction methods has emerged, spanning fully supervised learning with limited labels, semi-supervised learning that leverages unlabeled retired cells, weakly supervised approaches based on proxy or noisy labels, and unsupervised representation learning for clustering and anomaly detection. Each paradigm offers different trade-offs in terms of data requirement, interpretability, and deployment feasibility.

This section reviews these emerging technologies under a unified perspective. We first summarize the major routes for synthetic data generation (Section 4.2), followed by an integrated overview of health estimation approaches under various supervision settings (Section 4.3). A comparative evaluation highlighting accuracy, scalability, interpretability, and implementation complexity is provided in Section 4.4, before concluding with future opportunities toward reliable and scalable retirement-stage health prediction (Section 4.5).

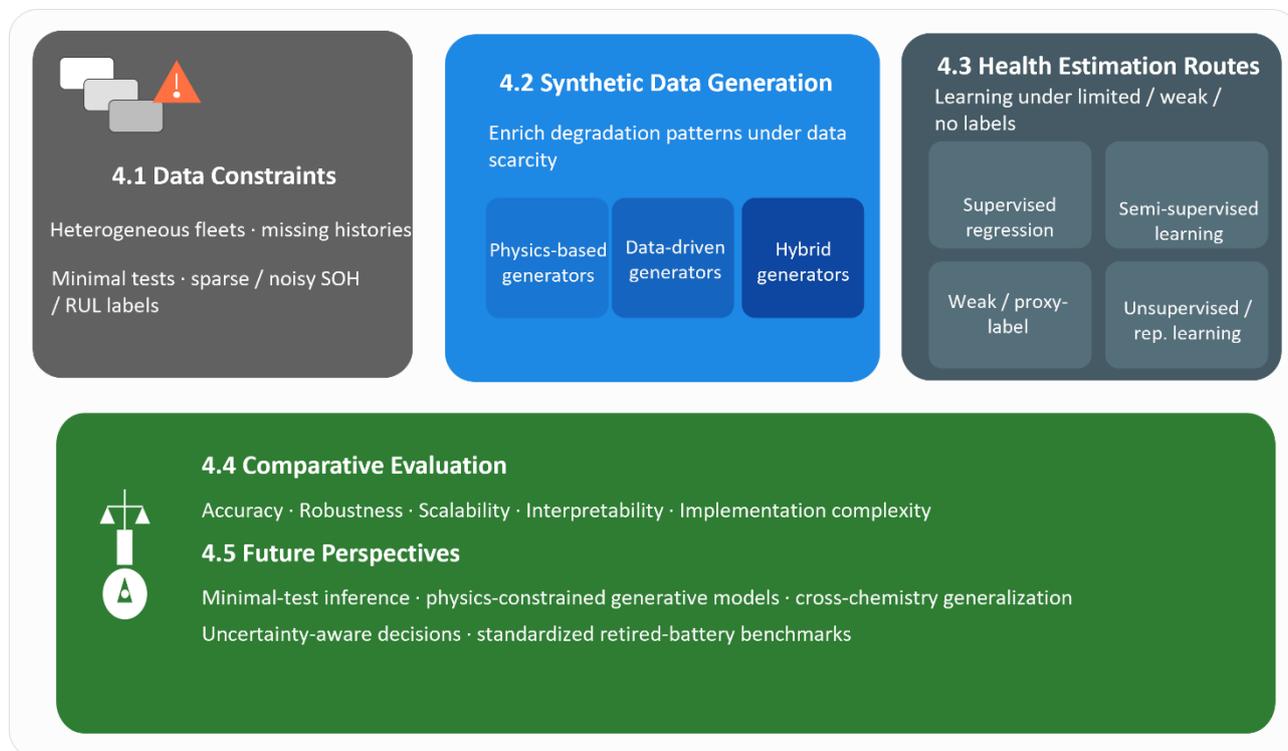

Fig.3 Retired Battery Health Estimation and Prediction Under Data Scarcity

## 4.1 Problem Setting

Retired lithium-ion batteries (R-LIBs) operate in a data environment profoundly different from that of well-instrumented laboratory-aged cells. Large variability in operational histories, usage conditions, and manufacturing sources leads to an intrinsically heterogeneous population, where degradation trajectories cannot be reconstructed from historical records. Retired batteries commonly originate from multiple cathode chemistries and manufacturers, with little or no shared operational data, making systematic characterization extremely challenging and often privacy-restricted due to ownership concerns [139]. This heterogeneity is a central barrier to achieving reliable estimation of SOH, RUL, or remaining capacity at retirement.

A major challenge arises from the absence of long-term cycling data, which prevents direct reconstruction of capacity fade, resistance growth, or mechanistic aging pathways. Zhao et al. [140] emphasize that many retired batteries enter second-life evaluation without accessible historical operating profiles, requiring diagnostic frameworks that operate independently of long-term data. Similarly, the work of Sulzer et al. [141] systematically reviews the limitations of field data, noting that end-use applications provide irregular, uncontrolled, and noisy measurements that significantly deviate from laboratory datasets traditionally used in aging-model development.

Furthermore, retired-battery assessment typically relies on minimal screening tests, such as short pulse injections, partial charging curves, or limited SOC sweeps, instead of full cycling or complete electrochemical impedance spectroscopy. Tao et al. [142] demonstrate that even large-scale datasets consisting of thousands of retired batteries require only single-pulse or short-range SOC tests, making measurement efficient but intrinsically low-information. Wang et al. [143] also highlight that field-collected data often have sparse sampling frequency, limited resolution, and inconsistent test protocols, severely restricting the reliability of impedance- or voltage-based diagnostics.

Another critical issue is label scarcity and label noise. Ground-truth SOH or capacity labels are rarely available in batch-level industrial screening due to the cost and time required for full capacity testing. Even when labels exist, they may be coarse, outdated, or affected by testing variability. Semi-supervised and weakly supervised works, such as Zhang et al. [144] and Wang et al. [145], show that practical datasets often contain a large proportion of unlabeled or weakly labeled samples, necessitating algorithms capable of learning from noisy or incomplete labels.

Compounding these challenges is the extreme population heterogeneity. Retired cells differ in cathode chemistry, form factor, thermal history, manufacturer design, and degradation pathway. Deep generative transfer learning frameworks, such as the study by Tao et al. [30], explicitly address the difficulty of aligning voltage-response dynamics across batteries that have undergone different aging mechanisms and SOC distributions. Large-format cells, as shown in Xie et al. [146], may also suffer from inhomogeneous internal degradation, including spatially localized lithium plating and deformation, which further complicates data-driven modeling.

Collectively, these constraints define a problem space characterized by irregular, incomplete, noisy, and highly heterogeneous measurements. Under such conditions, neither classical laboratory aging protocols nor purely supervised learning methods are sufficient for reliable retirement-stage health prediction. These limitations motivate the development of synthetic data generation, representation learning, and label-efficient predictive frameworks, which are examined in the following sections.

## 4.2   Synthetic Data Generation

The challenges outlined in Section 4.1, sparse measurements, missing historical records, label scarcity, and population-level heterogeneity, have motivated the increasing use of synthetic data generation to enrich the information space available for retired-battery health estimation. Unlike laboratory-aged cells, retired batteries rarely provide the full voltage, resistance, thermal, or impedance trajectories needed for robust SOH or RUL prediction. Synthetic samples can therefore serve as surrogates for incomplete degradation pathways, expand underrepresented aging modes, and improve model generalization across chemistries, manufacturers, and usage conditions. Recent studies have shown that augmenting limited screening data with physically consistent or statistically representative synthetic data significantly enhances predictive accuracy and robustness under random retirement conditions [30].

Approaches for synthetic data generation generally fall into three categories. Physics-based methods leverage electrochemical, thermal, or impedance models to emulate aging mechanisms with strong interpretability. Data-driven generative models, such as GANs, VAEs, and diffusion frameworks, learn degradation patterns directly from empirical data and can produce high-volume synthetic trajectories even when historical data are unavailable. Hybrid approaches combine physical constraints with representation learning to retain scientific plausibility while capturing the statistical diversity of retired-battery populations. These categories are reviewed in Sections 4.2.1–4.2.3.

### 4.2.1   Physics-Based Generation

Physics-based data generation aims to synthesize battery behaviors by explicitly modeling electrochemical, thermodynamic, and aging mechanisms. Compared with purely data-driven generators, these models preserve physical interpretability, ensure constraint consistency across operating conditions, and remain robust even when retired-battery data are sparse or noisy. Recent studies demonstrate that physical priors can be incorporated at multiple levels, from OCV reconstruction, impedance synthesis, and degradation-mechanism simulation to generative electrochemical modeling, thereby expanding the usable data space for downstream SOH/remaining-life estimation.

A representative line of work focuses on OCV–SOC curve reconstruction, which is essential for equilibrium-based health assessment but rarely available in field or retired-battery datasets. Wang et al. proposed a cloud-data-driven method that reconstructs complete OCV curves at different temperatures by combining analogy matching with an improved electrode-potential model, enabling continuous OCV updates during long-term operation [147]. Similarly, Xu et al. developed an online, training-free reconstruction framework that stitches discrete OCV fragments collected under uncontrolled conditions into a full OCV–SOC curve, demonstrating <5 mV reconstruction error across the battery's lifetime [148]. These studies highlight how physics-guided curve assembly can compensate for missing equilibrium data without requiring laboratory cycling.

Another important category involves electrochemical-impedance–based generative models. Doonyapisut et al. applied a variational autoencoder to explore large EIS datasets and map impedance spectra into a latent property space that aligns with physical circuit descriptors [149]. Such physics-consistent representations enable the generation of synthetic spectra and accelerate diagnostic model development. Likewise, Li et al. integrated an equivalent-circuit degradation model with synthetic electrochemical parameters to train a CNN for online parameter identification, addressing the limited availability of field-measurable impedance data in practical scenarios [150].

At a higher mechanistic level, several studies simulate battery aging and degradation trajectories by explicitly modeling SEI growth, lithium plating, charge-transfer resistance, and thermodynamic transitions. Jin et al. proposed a hybrid GAN framework that learns degradation trajectories under varied manufacturing and operating conditions, while embedding physics-derived argument data to alleviate data scarcity [151]. Franco et al. developed a physics-based simulation pipeline that generates high-fidelity synthetic electrodes and corresponding performance data, enabling machine-learning-assisted optimization of electrode designs [152]. More recently, Kim et al. introduced a physics-driven generative model (PGM) that samples electrochemical parameters and produces realistic synthetic cycling data, demonstrating high effectiveness in tasks such as internal short-circuit detection [153]. These models show that physically grounded parameter sampling can create synthetic retired-battery datasets that reflect realistic degradation modes even beyond the conditions present in available data.

Collectively, physics-based generation provides a principled path for enriching the scarce, inconsistent, and partially observable datasets typically encountered in retired-battery applications. The resulting synthetic data preserve mechanistic fidelity, support cross-chemistry generalization, and improve model robustness against the heterogeneous, noisy, and incomplete nature of field and screening data.

### 4.2.2  Data-Driven Generation

While physics-based generation provides mechanistic interpretability, it often struggles with heterogeneous retired-battery datasets where historical usage is unknown and degradation pathways vary widely. Data-driven generative models offer an alternative route by learning statistical representations directly from available field or laboratory observations, enabling the synthesis of realistic voltage curves, SOH trajectories, short-cycle snippets, or feature embeddings without explicit physical parameterization.

A major line of work employs GANs for augmenting SOH training data under limited or missing labels. Several studies demonstrate that GAN-based augmentation can effectively expand feature distributions of sparse datasets, reduce overfitting, and improve cross-battery adaptability. For example, one method applies a GAN with feature-mapping to align distributions across different batteries, improving SOH estimation accuracy with limited samples [154]. Other approaches integrate TimeGAN or recurrent structures to capture the temporal dependencies in multivariate degradation sequences, thereby generating synthetic long-term evolution under realistic operating variations [155]. Enhanced GAN variants such as W-DC-GAN-GP with Wasserstein distance and gradient penalty have also been proposed to improve stability and realism, and have been validated across multiple public datasets for SOH and SOC estimation [156]. In addition, CGANs have been shown to reconstruct missing or corrupted voltage curves and to generate informative samples for downstream estimators trained under various data-loss conditions [157].

Besides GANs, variational autoencoder (VAE)–based generators provide a probabilistic latent space for sampling diverse degradation patterns. VAE-driven augmentation has been used to generate representative battery aging data and enhance prediction performance of Transformer or neural sequence models, especially when dealing with noisy EV operational data [158]. Hybrid VAE–GAN frameworks have also been proposed to learn uncertainty-aware latent representations for RUL prediction, enabling probabilistic forecasting under real-world variability [159]. Extensions such as enhanced VAE (EVAE) combined with BiGRU or signal decomposition techniques further demonstrate the ability to embed EIS spectra, extract characteristic degradation components, and improve mapping to capacity or SOH trends [160].

A rapidly emerging direction involves generative transfer learning, where data generation is combined with domain-alignment techniques to handle heterogeneity across second-life or cross-condition datasets. Recent work proposes generating voltage dynamics across multiple SOC conditions and aligning domain gaps using correlation alignment (CORAL), achieving substantial reduction in data requirements for capacity estimation of heterogeneous second-life batteries [30]. Similar transfer-learning-based CGAN frameworks have been used for battery fault synthesis, enhancing classification performance under few-shot conditions by embedding fault priors during conditional retraining [161]. In addition, degradation-knowledge transfer combined with conditional time-series GANs enables synthetic data to support SOH modeling across different degradation modes and chemistries [162].

Overall, data-driven generative models offer a flexible and scalable solution for addressing the scarcity, incompleteness, and noise inherent to retired-battery datasets. However, their performance is sensitive to data representativeness, generative stability, and domain consistency, motivating the hybrid integration of physics priors and statistical generation, discussed in Section 4.2.3.

### 4.2.3 Hybrid Generation

Hybrid generative approaches aim to integrate mechanistic insights with data-driven learning to overcome the limitations of purely physics-based or purely statistical generation. While physics-based simulators can guarantee physical validity but often lack generalizability, data-driven generative models offer strong expressive power but may violate fundamental electrochemical principles or amplify noise, particularly problematic for heterogeneous retired-battery datasets. Hybrid methods seek to balance these strengths by embedding physical constraints, degradation mechanisms, or model-based priors into neural generative architectures.

One major category involves physics-informed synthetic data generation frameworks, where electrochemical models are used to generate high-rate or accelerated-aging voltage responses that enrich the training pool for data-driven estimators. For example, a pseudo-2D mechanistic aging model has been used to produce synthetic fast-charging voltage curves, which subsequently train neural networks capable of identifying lithium inventory loss and active material depletion under practical cycling conditions [163]. Such hybrid pipelines preserve physical interpretability while enabling scalable data augmentation beyond what is experimentally feasible.

A second line of research integrates physics-informed neural networks (PINN/PIML) with generative or representation-learning components. These methods embed degradation laws, equilibrium relationships, or partial differential equation (PDE) dynamics into the neural network structure or loss function. Hybrid PINN models have been applied for joint SOH/RUL estimation, short-term degradation path prediction, and multi-task battery health management, demonstrating improved generalization and physical consistency compared to traditional black-box networks [164], [165]. Physics-informed feature extraction—such as electrode-level physical variables, SEI-related descriptors, or incremental-capacity-derived constraints—has also been incorporated into generative or hybrid learning frameworks to enhance interpretability and reduce reliance on large labeled datasets [166], [167].

A third direction focuses on hybrid machine learning pipelines where physics-based models guide or constrain data-driven latent-space modeling. Examples include methods that combine physics-based long-term aging models (e.g., calendar and cycle aging equations) with LSTM architectures to learn degradation trajectories under varying environmental conditions [168], or hybrid degradation-model + GAN/AE frameworks that generate electrode-level or aging-mode–consistent synthetic data to support RUL prediction tasks [169]. These approaches ensure that generated samples follow realistic degradation pathways rather than arbitrary statistical variations.

Finally, hybridization has been extended to cross-domain or meta-learning frameworks where physical constraints inform adaptation under data scarcity. For instance, physics-informed loss functions derived from capacity-decay dynamics have been integrated into GRU–Transformer meta-learning architectures to enhance SOH estimation robustness across material systems and usage domains [170]. Similarly, hybrid physics-informed data generation combined with adaptive autoencoders has been shown to significantly improve RUL prediction when only limited labeled data are available, offering a promising pathway for retired-battery applications with incomplete or fragmented cycling histories [171].

Overall, hybrid physics–data-driven generative methods combine mechanistic validity with statistical flexibility, making them especially relevant for the retired-battery context, where heterogeneous aging, missing labels, and limited measurement granularity demand both physical plausibility and model adaptability. Remaining challenges include scalable physical constraint design, cross-chemistry generalization, and the development of unified evaluation benchmarks to compare hybrid generative pipelines.

## 4.3    Health Estimation Methods

Accurately assessing the health state of retired lithium-ion batteries requires analytical frameworks that can operate under data scarcity, heterogeneous usage histories, missing labels, and substantial measurement noise. Even with enhanced datasets obtained through screening tests and synthetic data generation, Section 4.2 has shown that the available information remains fundamentally imbalanced: some batteries provide only a few partial-cycle measurements, others include sparse SOH labels of uncertain quality, while a large population may contain no labels at all. As a result, the choice of learning paradigm, supervised, semi-supervised, weakly supervised, or unsupervised, becomes a central determinant of achievable prediction accuracy and robustness.

### 4.3.1    Supervised Learning

Supervised learning constitutes the most established paradigm for battery health estimation and therefore forms a natural starting point for discussing data-driven methods under retired-battery conditions. When a limited number of reliable SOH or RUL labels are available, typically obtained from screening tests, capacity checks, or partial-cycle measurements, supervised regression remains the default strategy for mapping health indicators to degradation states. However, in retired batteries these labels are often sparse, imprecise, or only partially representative of the true aging trajectory, which elevates the importance of label efficiency, structural robustness, and degradation-sensitive representation design. As a result, recent supervised approaches have evolved far beyond conventional "raw-signal regression," shifting toward physically guided and noise-tolerant formulations that maximize the value of scarce labeled samples while laying the foundation for the semi-supervised and weak-label paradigms discussed in subsequent sections.

Recent studies converge toward three overarching principles. First, supervised models increasingly rely on feature spaces that amplify degradation sensitivity while suppressing noise. Instead of learning directly from raw voltage or current trajectories, which demand large labeled datasets, methods transform measurements into structured representations such as aging-characteristic libraries, partial-recharge descriptors, or short-pulse features. One framework constructs a comprehensive library of over two hundred aging-related features and then applies embedded and filter-based selection to retain a compact subset for SOH regression across chemistries and operating conditions [53]. Other work engineers features from realistic random partial recharges, combining starting voltage, charge amount, and sliding-window statistics with a supervised selection pipeline to maintain SOH accuracy under mission-profile data [172]. For rapid screening of retired or second-life cells, short-term current-pulse tests are used to extract a few voltage knee-point features that are optimized via multi-objective search and then fed into support-vector regression, achieving a favorable trade-off between measurement cost and accuracy [173]. Indirect online SOH estimation adopts a similar philosophy by designing aggregate indicators such as constant-current drop capacity and maximum current rate during charging, which can be mapped to capacity loss via adaptive neural regressors under limited labeled data [174]. At system level, degradation-aware energy management and hybrid SOC/SOH observers likewise rely on handcrafted cycle-stress and specific-partial-capacity features coupled with relatively simple linear or Kalman-filter-based regressors [175], [176]. Synthetic yet interpretable degradation datasets generated from mechanistic models have also been exploited to train supervised random-forest classifiers on curve-derived features, enabling diagnosis of degradation modes with modest experimental labels [177].

Second, there is growing emphasis on regularization strategies and inductive biases that enforce physically meaningful behavior when labels are insufficient to constrain the model. Linear-in-parameters mappings from degradation features to life-loss metrics reduce the effective model degrees of freedom and help prevent overfitting when only a small number of retired cells are available [175]. Hybrid observers that couple adaptive Kalman filters with recursive least-squares updates on specific partial capacity explicitly model measurement noise and accumulation error, stabilizing SOC/SOH co-estimation over long operation periods [176]. In feature-based supervised estimators, monotonicity priors on capacity fade, smoothness constraints on predicted trajectories, and parameter sharing across cells with similar usage histories are increasingly integrated into the learning architecture, even when not always formalized as "physics-informed" models. These structural priors implicitly encode degradation knowledge and thereby compensate for incomplete or inconsistent labeling.

Third, supervised learning for retired batteries increasingly acknowledges that SOH/RUL targets may carry non-negligible measurement error and thus adopts robust label-handling mechanisms. When labels originate from fast capacity tests, DCIR-based screening, or approximate health indicators, noise-aware loss functions and label-smoothing strategies are introduced to reduce the influence of outliers. Prognosability-regularized frameworks exemplify this idea by training generative and predictive models jointly: synthetic SOH trajectories are constrained to remain smooth and degradation-consistent over time, and the supervised estimator is optimized under a loss that penalizes both prediction error and trend inconsistency, improving robustness with limited samples [178]. Beyond the battery field, multistage true-label distribution learning for mechanical fault diagnosis illustrates how distributional label modeling and adversarial loss designs can significantly enhance robustness to mislabeled data, offering transferable concepts for retired-battery SOH estimation under noisy labels [179].

In practice, many recent supervised pipelines combine these three principles, degradation-sensitive feature spaces, structured inductive biases, and label-robust loss designs, into unified frameworks. Even in their simplest form, such supervised models establish the baseline upon which semi-supervised, weakly supervised, and hybrid paradigms expand. In the context of retired batteries, where high-fidelity labels are scarce and expensive, supervised learning is thus shifting from "data-hungry regression on raw signals" to "physically guided, label-efficient inference," providing a stable reference against which more advanced methods in Sections 4.3.2–4.3.4 can be evaluated.

### 4.3.2    Semi-Supervised Learning

Semi-supervised learning (SSL) has become an indispensable direction for retired-battery health estimation because it directly addresses the structural imbalance between abundant unlabeled aging records and the very limited SOH/RUL labels available from screening tests. Recent advances show that effective SSL does not merely "add unlabeled data," but strategically transforms unlabeled or synthetic samples into reliable supervisory signals. Three methodological tendencies have gradually formed the backbone of SSL research in battery prognostics.

A major line of work strengthens pseudo-label generation by incorporating uncertainty modeling. This direction is motivated by the fact that naive pseudo-labeling easily propagates errors when labels are approximate, sparse, or inconsistent—conditions common in retired-battery datasets. Confidence-weighted pseudo-label refinement has therefore been increasingly adopted to suppress unreliable predictions and stabilize training, with studies demonstrating notable RMSE reductions when only a small fraction of samples are labeled [180]. Similar confidence-aware mechanisms have also been applied to RUL prediction, where pseudo-label reliability is explicitly weighted during optimization to reduce the impact of incorrect assignments [181]. These developments are consistent with broader PHM SSL research showing that manifold-regularized pseudo-label construction improves robustness when unlabeled data dominate [182].

A second group of semi-supervised approaches does not attempt to infer labels for unlabeled samples, but instead improves model robustness by enforcing representation consistency across perturbations, augmentations, or domain shifts. These methods rely exclusively on real unlabeled data, exploiting the fact that degradation trajectories exhibit stable geometric structure even without ground-truth SOH/RUL labels. For instance, consistency-regularized vision-transformer frameworks demonstrate that enforcing prediction invariance under multi-view or cross-distribution perturbations yields more stable degradation embeddings across chemistries and temperatures [183]. Likewise, dynamic-discharge SSL models combining BiGRU encoders with Gaussian process regularization maintain highly consistent representations even when calibration data are sparse [184]. Entropy-domain SSL methods further extend this idea by embedding multi-scale signal perturbation constraints directly into the feature space, improving label efficiency and reducing sensitivity to sampling noise [185]. Collectively, these techniques show that consistency priors provide an effective form of structural supervision when label scarcity, not model capacity, is the dominant limitation.

A conceptually distinct direction enhances SSL performance not by manipulating real unlabeled data, but by expanding the unlabeled data manifold itself through synthetic augmentation. In this paradigm, generated samples, produced via GANs, entropy-based feature construction, or physics-informed data generation, serve as auxiliary unlabeled inputs that broaden aging coverage before SSL is applied. Time-series GAN augmentation has been shown to enrich sparse battery datasets with realistic voltage–current–temperature trajectories, improving robustness in downstream SOH and SOC estimation [186]. Entropy-domain SSL studies report that synthetic-like multiscale features enhance manifold smoothness and strengthen pseudo-label refinement under strongly imbalanced real datasets [185]. Beyond batteries, industrial prognostics research demonstrates that embedding synthetic augmentation into online SSL improves adaptation to non-stationary conditions, a property highly relevant to heterogeneous retired-cell populations [187]. Synthetic-assisted SSL therefore forms a natural extension of the data-generation strategies discussed in Section 4.2, enabling richer unlabeled distributions without requiring additional physical testing.

Across these developments, SSL for retired cells is evolving from simple pseudo-label expansion toward a more structured exploitation of unlabeled and synthetic data, balancing confidence control, distribution alignment, and physically grounded augmentation. As retired-battery datasets remain inherently sparse, heterogeneous, and noisy, SSL provides a principled framework for leveraging available information without requiring large-scale relabeling. Future progress will depend on quantifying pseudo-label reliability, validating synthetic-data fidelity, and improving robustness to chemistry and condition shifts, key elements that define the next stage of SSL research for large-scale retired-battery screening and health prediction.

### 4.3.3    Weakly Supervised / Proxy-Label

Weakly supervised learning offers a practical route for health estimation when explicit SOH/RUL labels are scarce but proxy measurements, such as resistance evolution, pulse-based DCIR, EIS-derived features, or thermal-response indicators—are readily available from operational or screening tests. These proxy signals encode partial yet physically meaningful information about degradation, enabling models to learn health representations even in the absence of fully annotated cycle-life data. For example, multitask proxy-label frameworks have demonstrated that short charging segments can supervise deep models to predict SOH by learning auxiliary representations of degradation dynamics [188]. Likewise, relaxation-voltage features extracted using encoder–decoder architectures and clustered with Gaussian mixture models have been used as surrogate degradation stages, enabling coarse-to-fine SOH diagnosis without high-fidelity capacity labels [189].

When proxy labels are noisy or imprecise, uncertainty-aware learning becomes essential. Weak-label frameworks incorporating interval-based supervision explicitly treat SOH labels as bounded ranges rather than exact values, preventing overfitting to low-precision measurements and maintaining alignment with physical degradation constraints [188]. Self-supervised transformer pipelines fine-tuned with weak labels further demonstrate that proxy supervision can guide representation refinement while limiting the detrimental effects of mislabeled samples [145]. Complementing these strategies, noise-aware regression and confidence-weighted pseudo-label filtering can suppress unreliable supervision, significantly reducing error accumulation when labels originate from rapid screening or approximate SOH estimates [144].

Additional studies show that proxy labels can also be constructed from *domain-transformed representations*, offering alternative sources of weak supervision beyond traditional physical metrics. Gramian-angular-field–based transformations, multi-scale signal embeddings, and entropy-domain features have been used to generate auxiliary degradation cues that complement limited annotated data and improve downstream SOH estimation [145]. These representation-level proxies help models disentangle degradation patterns even when direct SOH annotations are sparse or noisy.

Overall, weakly supervised and proxy-label approaches occupy a middle ground between supervised and semi-supervised learning: proxy signals provide essential structural guidance, while model-based regularization mitigates their inherent imperfections. As screening-based measurements such as DCIR, relaxation voltage, and impedance parameters become increasingly available in industrial reuse pipelines, proxy-label learning is expected to play a growing role in scaling battery health estimation across large fleets of retired cells, bridging the gap between inexpensive operational tests and high-quality predictive performance.

### 4.3.4 Unsupervised Learning

Unsupervised learning has become a key pathway for retired-battery health estimation because it can extract degradation structure directly from field measurements without requiring explicit SOH/RUL labels. This paradigm is particularly attractive for large-scale second-life applications, where batteries arrive with heterogeneous histories and labels are absent, inconsistent, or prohibitively expensive to collect. Across recent studies, unsupervised methods converge around two complementary goals: discovering degradation manifolds and bridging domain gaps between diverse battery populations.

A first line of work leverages clustering-based health scoring to uncover latent structures that reflect degradation stages or emerging faults. By grouping cells using multi-feature embeddings, such as voltage statistics, temperature behavior, or impedance descriptors, unsupervised scoring systems can detect anomalous cells or early-stage thermal-runaway precursors without prior labeling [190]. In networked or distributed systems, clustering has been used not for fault detection but for predicting the longevity of node batteries, demonstrating how unsupervised SOH grouping can guide operational decisions at scale [191]. Cycle-aligned clustering frameworks further validate that grouping based on cross-cycle health factors enhances RUL prediction when labeled samples are sparse, as cluster-level structure provides an additional source of regularization for downstream regressors [192]. These examples collectively show that clustering transforms unlabeled field data into physically meaningful representations that support both diagnostics and lifetime forecasting.

A second line of research tackles the domain-shift challenge inherent in retired batteries, namely, that training and deployment data often come from different chemistries, usage patterns, or environments. Unsupervised domain adaptation approaches learn transferable feature spaces in which distribution mismatches are minimized, enabling models trained on one battery population to generalize to another. Recent work has demonstrated that geometric metrics between feature subspaces can be optimized to reduce cross-domain discrepancies and thus stabilize SOH estimation without labels in the target domain [193]. Deep subdomain adaptation networks extend this idea by combining CNN encoders, LSTM temporal extractors, and attention mechanisms to align multi-source data at a finer granularity [194]. In parallel, domain-adversarial transfer learning frameworks employ adversarial objectives or maximum-mean-discrepancy criteria to learn domain-invariant degradation features, yielding notable improvements in cross-battery SOH prediction [195], [196]. Together, these studies highlight that representation alignment is crucial when retired batteries originate from mixed fleets and fragmented operational histories.

Beyond clustering and domain adaptation, an emerging direction centers on self-supervised representation learning, which extracts degradation-sensitive embeddings from unlabeled cycling trajectories before fine-tuning with a minimal number of labeled samples. Self-supervised encoders trained on partial voltage curves or dynamic charging data have shown strong gains in label efficiency and robustness to noisy operating conditions [197]. Meanwhile, unsupervised pattern-discovery methods compress large volumes of vehicle load-profile data into representative patterns, enabling efficient LSTM training even when raw time series are high-volume and highly variable [198]. These approaches demonstrate that powerful representations can be learned entirely from unlabeled signals, reducing dependence on costly full-cycle characterization.

Overall, unsupervised and representation-learning methods provide a complementary foundation to supervised and semi-supervised approaches by mapping heterogeneous retired-battery data into structured, degradation-aware feature spaces. Clustering reveals intrinsic health structure, domain-adaptation mitigates distribution mismatch, and self-supervised learning distills degradation signatures from raw sequences. As the diversity and scale of retired-battery inventories grow, these unsupervised paradigms are expected to play an increasingly central role in establishing consistent, transferable, and label-efficient health estimation frameworks.

## 4.4    Comparative Evaluation

Battery health estimation under retired-battery scenarios involves multiple modeling routes—from supervised learning to semi-supervised, weakly supervised, and unsupervised paradigms. While each route addresses different forms of data scarcity, noise, or heterogeneity, a cross-method comparison highlights several overarching trends that shape their practical utility. A fundamental observation is that performance gains increasingly depend on how well each route reconciles the mismatch between limited ground-truth information and high variability in aging patterns. Supervised learning typically delivers the highest accuracy when reliable labels exist, but its robustness deteriorates rapidly under label noise, profile irregularity, or mixed chemistries. Semi-supervised and self-supervised approaches moderate this decline by leveraging unlabeled structure—consistency priors, pseudo-labels, or synthetic augmentation—but remain sensitive to domain shifts if the unlabeled data fail to span representative degradation trajectories. Unsupervised clustering and domain adaptation offer strong adaptability across chemistries and usage conditions, though their predictive power is generally constrained by the lack of explicit supervisory signals.

A second key difference lies in scalability to heterogeneous retired-battery populations. Supervised pipelines often require chemistry- or manufacturer-specific calibration, as even small differences in electrode formulation or operating history can bias the learned degradation manifold. Methods that incorporate latent-space alignment, distribution matching, or domain-adversarial regularization achieve better transferability across chemistries, particularly when trained with mixed-population datasets. Semi-supervised strategies using synthetic data can further expand the degradation manifold, improving generalization in scenarios where only partial aging ranges are captured by real data. By contrast, unsupervised clustering and subspace-alignment approaches inherently adapt to cross-dataset variability, making them attractive for early screening or pre-classification of diverse retired-battery inventories.

Interpretability and safety considerations also differentiate modeling routes. Physics-guided supervised methods—e.g., using impedance-derived features, plateau characteristics, or recovery-based indicators—tend to yield explicit and traceable degradation signatures, which is valuable for second-life qualification. Weakly supervised approaches preserve interpretability when proxy labels reflect physical mechanisms such as resistance growth or thermal response trends. In contrast, purely unsupervised and domain-adversarial frameworks sacrifice some transparency in exchange for flexibility, raising challenges for safety-critical deployment unless complemented with post-hoc diagnostics or fault-signature reconstruction. As real-world reuse increasingly requires explainability for regulatory approval and risk assessment, methods that embed physical structure or support interpretable feature spaces retain a practical advantage.

Finally, computational efficiency and implementation complexity vary substantially across routes. Lightweight supervised models or proxy-label regression require minimal tuning and are well suited for industrial sorting lines. SSL frameworks with synthetic augmentation introduce additional training stages but maintain manageable complexity. Domain-adversarial and deep subdomain-adaptation models offer strong cross-population transfer but often require large networks, adversarial optimization, and extensive hyperparameter control. Unsupervised clustering and similarity-metric methods remain comparatively simple but rely heavily on high-quality feature engineering. In real deployment, the preferred approach often reflects a balance between hardware constraints, data availability, and the required accuracy–robustness–explainability trade-off.

Overall, a comparative view reveals that no single modeling route is universally superior; rather, each fills a distinct role along the spectrum of data availability and operational constraints. For retired-battery applications—where labels are sparse, chemistries are diverse, and safety is paramount—hybrid strategies that integrate physics-guided features, representation learning, and domain-robust adaptation appear most promising, forming a flexible foundation for scalable and reliable health estimation in second-life ecosystems.

# Table 2 — Comparative Evaluation of Modeling Routes for Battery Health Estimation

| Modeling Route | Typical Input Requirement | Label Dependence | Accuracy (Under Clean Data) | Robustness (Noise / Mixed Chemistries) | Interpretability | Scalability (Field / Second-Life) | Computational Complexity | Practical Cost & Deployment Suitability |
|---|---|---|---|---|---|---|---|---|
| *Supervised Learning* | Full charge/discharge or carefully designed diagnostic profiles | High (precise SOH/RUL required) | ★★★★☆ (best when labels reliable) | ★★☆☆☆ (performance drops under domain shift or label noise) | ★★★★☆ (feature-based or physics-guided models interpretable) | ★★☆☆☆ (chemistry-specific calibration often required) | ★★☆☆☆ (lightweight to moderate) | High cost (requires capacity tests or controlled lab aging data) |
| *Semi-Supervised Learning* | Mixed labeled + unlabeled charging snippets, partial cycling data | Medium | ★★★★☆ (near-supervised with enough unlabeled structure) | ★★★☆☆ (depends on pseudo-label reliability and domain mismatch) | ★★★☆☆ (latent-space representations less transparent) | ★★★★☆ (better transfer across populations using unlabeled structure) | ★★★☆☆ (consistency training, SSL pipelines) | Medium cost (reduces labeling demand; requires data preprocessing) |
| *Weakly Supervised (Proxy-Label) Learning* | DCIR, EIS, thermal rise, rest-voltage, or other fast measurements | Low–Medium (proxy quality is key) | ★★★☆☆ (limited by proxy fidelity) | ★★★★☆ (physically grounded proxies increase robustness) | ★★★★☆ (proxy labels reflect physical mechanisms) | ★★★★★ (highly scalable for industrial sorting & retired batteries) | ★★☆☆☆ (simple regression/classific ation) | Low cost (fast tests; no capacity measurements needed) |
| *Unsupervised / Clustering / Domain Adaptation* | Raw profiles, relaxation voltages, partial snippets, heterogeneous datasets | None | ★★☆☆☆ (no direct SOH target) | ★★★★☆ (strong adaptability; domain-invariant representations) | ★★☆☆☆ (latent clusters hard to interpret) | ★★★★★ (excellent for diverse chemistries and mixed field data) | ★★★★☆ (adversarial training, subspace alignment, similarity metrics) | Very low labeling cost, but higher training overhead |
| *Self-Supervised (Contrastive / Pretext Tasks)* | Large-scale unlabeled time-series; partial CV/CC segments | Low | ★★★★☆ (strong representations when pretraining succeeds) | ★★★★☆ (pretext tasks enhance invariance to noise and profiles) | ★★★☆☆ (latent features; depends on design) | ★★★★☆ (can generalize across similar operational regimes) | ★★★★☆ (requires pretraining + fine-tuning pipeline) | Medium cost (large unlabeled data required) |

## 4.5    Future Perspectives

Looking ahead, the development of reliable health prediction frameworks for retired lithium-ion batteries will depend on unifying sparse physical measurements with scalable data-driven inference. Since full diagnostic cycling is infeasible at end-of-life sorting stages, future methods will increasingly rely on minimal and low-cost tests—such as short current pulses, partial charge segments, rest-voltage windows, impedance snippets, or thermal-rise responses—combined with representation learning capable of extracting degradation-sensitive features from these incomplete signals. Such "measure minimally, infer maximally" pipelines have the potential to shorten evaluation time by orders of magnitude while retaining actionable accuracy for reuse decisions.

Another important direction concerns the incorporation of physical structure into generative and augmentation models. While synthetic data already improve model robustness under limited labels, unconstrained generators may inadvertently violate electrochemical principles or produce unrealistic degradation trajectories. Future advances are expected to embed constraints derived from thermodynamics, impedance evolution, phase-transition behavior, or monotonic capacity fade directly within generative architectures. These physically guided models can ensure that augmented datasets remain credible, interpretable, and aligned with real aging mechanisms, thereby enabling safer and more trustworthy downstream learning.

Generalization across chemistries, formats, and real-world operating conditions also represents a central challenge. Retired batteries enter second-life markets from highly diverse sources, NCM, LFP, NCA, pouch cells, cylindrical formats, EV duty cycles, residential storage, and more. To handle this heterogeneity, future research will require domain-invariant representation learning capable of transferring degradation knowledge across distinct electrochemical signatures. Approaches combining large-scale self-supervised pretraining, domain-adversarial alignment, and cross-population contrastive objectives may provide the foundation for building models that maintain predictive fidelity even when test batteries exhibit behavior not seen during training.

Because retired-battery decisions affect safety, warranty valuation, and system-level risk, uncertainty quantification will become increasingly important. Instead of single-point SOH predictions, next-generation models must provide calibrated confidence intervals that reflect measurement noise, proxy-label ambiguity, and distribution shift. Probabilistic regression, evidential deep learning, Bayesian ensembles, and conformal prediction are likely to play growing roles in enabling risk-aware decision thresholds for pack sorting, module matching, and second-life deployment.

Finally, progress in this field will depend on the establishment of standardized benchmarks tailored specifically to retired batteries. Current datasets vary widely in test conditions, aging protocols, chemistries, and labeling practices, making cross-study comparison difficult. The creation of shared evaluation protocols—covering minimal-test inputs, proxy-label definitions, domain-shift scenarios, and performance metrics beyond accuracy—would provide a common basis for evaluating robustness, interpretability, and computational efficiency. The emergence of such benchmarks is essential for transforming research innovations into practical industrial tools for large-scale, safe, and economically viable battery repurposing.

# 5 Conclusion

Retired-battery health assessment presents a uniquely challenging problem: measurements are sparse and heterogeneous, ground-truth labels are limited or imprecise, aging trajectories vary widely across chemistries and prior usage histories, and safety requirements constrain the range of allowable tests. This review has synthesized the diverse modeling routes developed to address these constraints, spanning supervised, semi-supervised, weakly supervised, and unsupervised paradigms, and has clarified how each contributes distinct capabilities for extracting degradation information from incomplete or noisy data. By examining data-generation strategies, representation-learning techniques, proxy-label construction, and domain-adaptation frameworks, we highlight the emerging convergence toward hybrid pipelines that fuse minimal physical tests with scalable learning-based inference.

Across methods, several persistent challenges stand out, including the need for models that remain robust under domain shift, can leverage large unlabeled corpora without propagating noise, and produce uncertainty-aware predictions suitable for safety-critical reuse decisions. At the same time, opportunities are rapidly expanding: generative models increasingly integrate electrochemical principles, self-supervised pretraining yields transferable degradation representations, and unsupervised alignment techniques allow models to generalize across chemistries and operating regimes. Together, these developments point toward a new generation of retirement-stage health estimation frameworks that are both data-efficient and physically grounded.

As second-life applications scale globally, reliable and standardized diagnostic methodologies will be essential for ensuring safety, performance, and economic viability. Progress will depend not only on algorithmic advances but also on the creation of common benchmarks, shared datasets, and reproducible evaluation protocols tailored to the realities of retired-battery populations. By consolidating the current landscape and identifying unifying trends, this review aims to support the development of robust, interpretable, and industry-ready health estimation approaches that can unlock the full value of lithium-ion batteries beyond their first life.

# Reference

[1] G. Harper *et al.*, "Recycling lithium-ion batteries from electric vehicles," *Nature*, vol. 575, no. 7781, pp. 75–86, Nov. 2019

[2] L. Colarullo and J. Thakur, "Second-life EV batteries for stationary storage applications in local energy communities," *Renew. Sustain. Energy Rev.*, vol. 169, p. 112913, Nov. 2022

[3] H. Wang, M. Rasheed, R. Hassan, M. Kamel, S. Tong, and R. Zane, "Life-extended active battery control for energy storage using electric vehicle retired batteries," *IEEE Trans. Power Electron.*, vol. 38, no. 6, pp. 6801–6805, June 2023

[4] J. Zhu *et al.*, "End-of-life or second-life options for retired electric vehicle batteries," *Cell Rep. Phys. Sci.*, vol. 2, no. 8, p. 100537, Aug. 2021

[5] X. Cui, M. A. Khan, G. Pozzato, S. Singh, R. Sharma, and S. Onori, "Taking second-life batteries from exhausted to empowered using experiments, data analysis, and health estimation," *Cell Rep. Phys. Sci.*, vol. 5, no. 5, p. 101941, May 2024

[6] Y. Zhang *et al.*, "Performance assessment of retired EV battery modules for echelon use," *Energy*, vol. 193, p. 116555, Feb. 2020

[7] T. Wang, Y. Jiang, L. Kang, and Y. Liu, "Determination of retirement points by using a multi-objective optimization to compromise the first and second life of electric vehicle batteries," *J. Clean. Prod.*, vol. 275, p. 123128, Dec. 2020

[8] Q. Dong, S. Liang, J. Li, H. C. Kim, W. Shen, and T. J. Wallington, "Cost, energy, and carbon footprint benefits of second-life electric vehicle battery use," *iScience*, vol. 26, no. 7, p. 107195, July 2023

[9] M. H. S. M. Haram, J. W. Lee, G. Ramasamy, E. E. Ngu, S. P. Thiagarajah, and Y. H. Lee, "Feasibility of utilising second life EV batteries: Applications, lifespan, economics, environmental impact, assessment, and challenges," *Alex. Eng. J.*, vol. 60, no. 5, pp. 4517–4536, Oct. 2021

[10] E. Martinez-Laserna *et al.*, "Battery second life: Hype, hope or reality? A critical review of the state of the art," *Renew. Sustain. Energy Rev.*, vol. 93, pp. 701–718, Oct. 2018

[11] J. Li, S. He, Q. Yang, Z. Wei, Y. Li, and H. He, "A comprehensive review of second life batteries toward sustainable mechanisms: Potential, challenges, and future prospects," *IEEE Trans. Transp. Electrification*, vol. 9, no. 4, pp. 4824–4845, Dec. 2023

[12] J. Wu, J. Wang, M. Lin, and J. Meng, "Retired battery capacity screening based on deep learning with embedded feature smoothing under massive imbalanced data," *Energy*, vol. 318, p. 134761, Mar. 2025

[13] Md. T. Sarker, M. H. S. M. Haram, S. J. Shern, G. Ramasamy, and F. Al Farid, "Second-life electric vehicle batteries for home photovoltaic systems: Transforming energy storage and sustainability," *Energies*, vol. 17, no. 10, p. 2345, May 2024

[14] Y. Chen *et al.*, "A review of lithium-ion battery safety concerns: The issues, strategies, and testing standards," *J. Energy Chem.*, vol. 59, pp. 83–99, Aug. 2021

[15] J. Liu, L. Zhou, Y. Zhang, J. Wang, and Z. Wang, "Aging behavior and mechanisms of lithium-ion battery under multi-aging path," *J. Clean. Prod.*, vol. 423, p. 138678, Oct. 2023

[16] Y. Qian *et al.*, "Influence of electrolyte additives on the cathode electrolyte interphase (CEI) formation on LiNi1/3Mn1/3Co1/3O2 in half cells with li metal counter electrode," *J. Power Sources*, vol. 329, pp. 31–40, Oct. 2016

[17] S. Xie *et al.*, "Influence of cycling aging and ambient pressure on the thermal safety features of lithium-ion battery," *J. Power Sources*, vol. 448, p. 227425, Feb. 2020

[18] D. Ouyang, J. Weng, M. Chen, J. Wang, and Z. Wang, "Electrochemical and thermal features of aging lithium-ion batteries cycled at various current rates," *J. Loss Prev. Process Ind.*, vol. 85, p. 105156, Oct. 2023

[19]  R. Xiong, Y. Pan, W. Shen, H. Li, and F. Sun, "Lithium-ion battery aging mechanisms and diagnosis method for automotive applications: Recent advances and perspectives," *Renew. Sustain. Energy Rev.*, vol. 131, p. 110048, Oct. 2020

[20]  M. Dubarry *et al.*, "Identifying battery aging mechanisms in large format li ion cells," *J. Power Sources*, vol. 196, no. 7, pp. 3420–3425, Apr. 2011

[21]  Z. Wang, Q. Zhao, S. Wang, Y. Song, B. Shi, and J. He, "Aging and post-aging thermal safety of lithium-ion batteries under complex operating conditions: A comprehensive review," *J. Power Sources*, vol. 623, p. 235453, Dec. 2024

[22]  T. Gao *et al.*, "Effect of aging temperature on thermal stability of lithium-ion batteries: Part a – high-temperature aging," *Renew. Energy*, vol. 203, pp. 592–600, Feb. 2023

[23]  R. Li *et al.*, "Accelerated aging of lithium-ion batteries: Bridging battery aging analysis and operational lifetime prediction," *Sci. Bull.*, vol. 68, no. 23, pp. 3055–3079, Dec. 2023

[24]  Y. Wu *et al.*, "Optimal battery thermal management for electric vehicles with battery degradation minimization," *Appl. Energy*, vol. 353, p. 122090, Jan. 2024

[25]  X. Hu *et al.*, "A review of second-life lithium-ion batteries for stationary energy storage applications," *Proc. IEEE*, vol. 110, no. 6, pp. 735–753, June 2022

[26]  W. Kong *et al.*, "Residual capacity estimation and consistency sorting of retired lithium batteries in cascade utilization process: A review," *Green Manuf. Open*, vol. 3, no. 1, Jan. 2025 Accessed: Nov. 16, 2025. [Online]. Available: https://www.oaepublish.com/articles/gmo.2024.111301

[27]  Y. Ni *et al.*, "Accurate estimation of residual capacity for large-scale retired-LFP batteries with multiple aging pathways," *J. Energy Storage*, vol. 127, p. 117111, Aug. 2025

[28]  R. Guo, F. Wang, M. Akbar Rhamdhani, Y. Xu, and W. Shen, "Managing the surge: A comprehensive review of the entire disposal framework for retired lithium-ion batteries from electric vehicles," *J. Energy Chem.*, vol. 92, pp. 648–680, May 2024

[29]  Q. Zhang, X. Li, Z. Du, and Q. Liao, "Aging performance characterization and state-of-health assessment of retired lithium-ion battery modules," *J. Energy Storage*, vol. 40, p. 102743, Aug. 2021

[30]  S. Tao *et al.*, "Immediate remaining capacity estimation of heterogeneous second-life lithium-ion batteries *via* deep generative transfer learning," *Energy Environ. Sci.*, vol. 18, no. 15, pp. 7413–7426, 2025

[31]  A. Garg, L. Yun, L. Gao, and D. B. Putungan, "Development of recycling strategy for large stacked systems: Experimental and machine learning approach to form reuse battery packs for secondary applications," *J. Clean. Prod.*, vol. 275, p. 124152, Dec. 2020

[32]  M. Rasheed *et al.*, "Active reconditioning of retired lithium-ion battery packs from electric vehicles for second-life applications," *IEEE J. Emerg. Sel. Top. Power Electron.*, vol. 12, no. 1, pp. 388–404, Feb. 2024

[33]  Y. Wang, H. Huang, and H. Wang, "A new method for fast state of charge estimation using retired battery parameters," *J. Energy Storage*, vol. 55, p. 105621, Nov. 2022

[34]  H. Hong and Y. Zhu, "Evaluation of lithium battery cycle aging based on temperature increase during charging"

[35]  S.-S. Yun and S.-C. Kee, "Effect of capacity variation in series-connected batteries on aging," *Batteries*, vol. 9, no. 1, p. 22, Dec. 2022

[36]  Y. Wang, H. Huang, and H. Wang, "Rapid-regroup strategy for retired batteries based on short-time dynamic voltage and electrochemical impedance spectroscopy," *J. Energy Storage*, vol. 63, p. 107102, July 2023

[37]  Z. Lyu, Y. Zhang, G. Wang, and R. Gao, "A semiparametric clustering method for the screening of retired li-ion batteries from electric vehicles," *J. Energy Storage*, vol. 63, p. 107030, July 2023

[38]  X. Lai, C. Deng, J. Li, Z. Zhu, X. Han, and Y. Zheng, "Rapid sorting and regrouping of retired lithium-ion battery modules for echelon utilization based on partial charging curves," *IEEE Trans. Veh. Technol.*, vol. 70, no. 2, pp. 1246–1254, Feb. 2021

[39]    S. Duan, Z. Yu, J. Li, Z. Zhao, and H. Liu, "Rapid screening for retired batteries based on lithium-ion battery IC curve prediction," *World Electr. Veh. J.*, vol. 15, no. 10, p. 451, Oct. 2024

[40]    M. Lin, Z. Xu, G. Zheng, J. Meng, W. Wang, and J. Wu, "Retired lithium-ion batteries screening based on partial discharge curves and an improved CrossFormer," *IEEE Trans. Transp. Electrification*, vol. 11, no. 4, pp. 10239–10249, Aug. 2025

[41]    T. Liu, X. Chen, Q. Peng, J. Peng, and J. Meng, "An enhanced sorting method for retired battery with feature selection and multiple clustering," *J. Energy Storage*, vol. 87, p. 111422, May 2024

[42]    X. Liu, Q. Tang, Y. Feng, M. Lin, J. Meng, and J. Wu, "Fast sorting method of retired batteries based on multi-feature extraction from partial charging segment," *Appl. Energy*, vol. 351, p. 121930, Dec. 2023

[43]    R. Ma *et al.*, "Pathway decisions for reuse and recycling of retired lithium-ion batteries considering economic and environmental functions," *Nat. Commun.*, vol. 15, no. 1, p. 7641, Sept. 2024

[44]    T. Montes, M. Etxandi-Santolaya, J. Eichman, V. J. Ferreira, L. Trilla, and C. Corchero, "Procedure for assessing the suitability of battery second life applications after EV first life," *Batteries*, vol. 8, no. 9, p. 122, Sept. 2022

[45]    Z. Zhou *et al.*, "A fast screening framework for second-life batteries based on an improved bisecting K-means algorithm combined with fast pulse test," *J. Energy Storage*, vol. 31, p. 101739, Oct. 2020

[46]    Y. Zhang, Z. Zhou, Y. Kang, C. Zhang, and B. Duan, "A quick screening approach based on fuzzy C-means algorithm for the second usage of retired lithium-ion batteries," *IEEE Trans. Transp. Electrification*, vol. 7, no. 2, pp. 474–484, June 2021

[47]    A. Weng, E. Dufek, and A. Stefanopoulou, "Battery passports for promoting electric vehicle resale and repurposing," *Joule*, vol. 7, no. 5, pp. 837–842, May 2023

[48]    C. A. Rufino Júnior *et al.*, "Towards to battery digital passport: Reviewing regulations and standards for second-life batteries," *Batteries*, vol. 10, no. 4, p. 115, Mar. 2024

[49]    G. Seo *et al.*, "Rapid determination of lithium-ion battery degradation: High C-rate LAM and calculated limiting LLI," *J. Energy Chem.*, vol. 67, pp. 663–671, Apr. 2022

[50]    C. Wang *et al.*, "State estimation and aging mechanism of 2nd life lithium-ion batteries: Non-destructive and postmortem combined analysis," *Electrochimica Acta*, vol. 443, p. 141996, Mar. 2023

[51]    R. Xiong, P. Wang, Y. Jia, W. Shen, and F. Sun, "Multi-factor aging in lithium iron phosphate batteries: Mechanisms and insights," *Appl. Energy*, vol. 382, p. 125250, Mar. 2025

[52]    G. Li, B. Li, C. Li, and S. Wang, "State-of-health rapid estimation for lithium-ion battery based on an interpretable stacking ensemble model with short-term voltage profiles," *Energy*, vol. 263, p. 126064, Jan. 2023

[53]    J. Wang *et al.*, "A novel aging characteristics-based feature engineering for battery state of health estimation," *Energy*, vol. 273, p. 127169, June 2023

[54]    X. Lai *et al.*, "Voltage profile reconstruction and state of health estimation for lithium-ion batteries under dynamic working conditions," *Energy*, vol. 282, p. 128971, Nov. 2023

[55]    S. Peng, J. Zhu, T. Wu, A. Tang, J. Kan, and M. Pecht, "SOH early prediction of lithium-ion batteries based on voltage interval selection and features fusion," *Energy*, vol. 308, p. 132993, Nov. 2024

[56]    B.-R. Chen, C. M. Walker, S. Kim, M. R. Kunz, T. R. Tanim, and E. J. Dufek, "Battery aging mode identification across NMC compositions and designs using machine learning," *Joule*, vol. 6, no. 12, pp. 2776–2793, Dec. 2022

[57]    X. Han, M. Ouyang, L. Lu, J. Li, Y. Zheng, and Z. Li, "A comparative study of commercial lithium ion battery cycle life in electrical vehicle: Aging mechanism identification," *J. Power Sources*, vol. 251, pp. 38–54, Apr. 2014

[58]    G. Wang, N. Cui, C. Li, Z. Cui, and H. Yuan, "A state-of-health estimation method based on incremental capacity analysis for li-ion battery considering charging/discharging rate," *J. Energy Storage*, vol. 73, p. 109010, Dec. 2023

[59]    R. Xiong, P. Wang, Y. Jia, W. Shen, and F. Sun, "Multi-factor aging in lithium iron phosphate batteries: Mechanisms and insights," *Appl. Energy*, vol. 382, p. 125250, Mar. 2025

[60] F. Wang *et al.*, "Capacity prediction of lithium-ion batteries with fusing aging information," *Energy*, vol. 293, p. 130743, Apr. 2024

[61] J. Wen, X. Chen, X. Li, and Y. Li, "SOH prediction of lithium battery based on IC curve feature and BP neural network," *Energy*, vol. 261, p. 125234, Dec. 2022

[62] F. Wang, Z. Zhai, Z. Zhao, Y. Di, and X. Chen, "Physics-informed neural network for lithium-ion battery degradation stable modeling and prognosis," *Nat. Commun.*, vol. 15, no. 1, p. 4332, May 2024

[63] Q. Zhang, X. Li, Z. Du, and Q. Liao, "Aging performance characterization and state-of-health assessment of retired lithium-ion battery modules," *J. Energy Storage*, vol. 40, p. 102743, Aug. 2021

[64] L. Von Kolzenberg, J. Stadler, J. Fath, M. Ecker, B. Horstmann, and A. Latz, "A four parameter model for the solid-electrolyte interphase to predict battery aging during operation," *J. Power Sources*, vol. 539, p. 231560, Aug. 2022

[65] B. Nie, Y. Dong, L. Chang, and L. Zeng, "Experimental investigation of cycling aging effects on degradation and thermal runaway characteristics of 18650 lithium-ion batteries," *J. Energy Storage*, vol. 134, p. 118171, Oct. 2025

[66] S. Zhang, X. Guo, X. Dou, and X. Zhang, "A rapid online calculation method for state of health of lithium-ion battery based on coulomb counting method and differential voltage analysis," *J. Power Sources*, vol. 479, p. 228740, Dec. 2020

[67] P. Yong, F. Guo, and Z. Yang, "An age-dependent battery energy storage degradation model for power system operations," *IEEE Trans. Power Syst.*, vol. 40, no. 1, pp. 1188–1191, Jan. 2025

[68] B. Larvaron, "Chained gaussian processes with derivative information to forecast battery health degradation," *J. Energy Storage*, 2023

[69] D. Clerici, F. Pistorio, and A. Somà, "Aging diagnostics in lithium-ion batteries with differential mechanical measurements," *Appl. Energy*, vol. 386, p. 125524, May 2025

[70] B. Pan *et al.*, "Aging mechanism diagnosis of lithium ion battery by open circuit voltage analysis," *Electrochimica Acta*, vol. 362, p. 137101, Dec. 2020

[71] A. Bavand, S. A. Khajehoddin, M. Ardakani, and A. Tabesh, "Online estimations of li-ion battery SOC and SOH applicable to partial charge/discharge," *IEEE Trans. Transp. Electrification*, vol. 8, no. 3, pp. 3673–3685, Sept. 2022

[72] S. Barcellona, S. Colnago, and L. Codecasa, "Combined effect of cycle aging and temperature on the variation of the open-circuit voltage of lithium cobalt oxide batteries," *J. Power Sources*, vol. 660, p. 238479, Dec. 2025

[73] Z. Cui, N. Cui, C. Li, J. Lu, and C. Zhang, "Online identification and reconstruction of open-circuit voltage for capacity and electrode aging estimation of lithium-ion batteries," *IEEE Trans. Ind. Electron.*, vol. 70, no. 5, pp. 4716–4726, May 2023

[74] G. Fan and X. Zhang, "Battery capacity estimation using 10-second relaxation voltage and a convolutional neural network," *Appl. Energy*, vol. 330, p. 120308, Jan. 2023

[75] R. Wang, J. Li, X. Wang, S. Wang, and M. Pecht, "Deep learning model for state of health estimation of lithium batteries based on relaxation voltage," *J. Energy Storage*, vol. 79, p. 110189, Feb. 2024

[76] A. Tang *et al.*, "Deep learning driven battery voltage-capacity curve prediction utilizing short-term relaxation voltage," *eTransportation*, vol. 22, p. 100378, Dec. 2024

[77] A. Reiter *et al.*, "Exploring the effects of aging, temperature and hysteresis on the entropy variation of lithium-ion batteries," 2025, *SSRN*. Accessed: Nov. 26, 2025. [Online]. Available: https://www.ssrn.com/abstract=5142673

[78] P. Yu, S. Wang, C. Yu, W. Shi, and B. Li, "Study of hysteresis voltage state dependence in lithium-ion battery and a novel asymmetric hysteresis modeling," *J. Energy Storage*, vol. 51, p. 104492, July 2022

[79] J. Schmitt, I. Horstkötter, and B. Bäker, "Data efficient open circuit voltage hysteresis modelling – transfer fitting the trajectory correction hysteresis (TCH) model from SOH-to-SOH and different li-ion cell chemistries," *J.*

*Power Sources Adv.*, vol. 27, p. 100146, June 2024

[80]    P. Svens, A. J. Smith, J. Groot, M. J. Lacey, G. Lindbergh, and R. W. Lindstrom, "Evaluating performance and cycle life improvements in the latest generations of prismatic lithium-ion batteries," *IEEE Trans. Transp. Electrification*, vol. 8, no. 3, pp. 3696–3706, Sept. 2022

[81]    K. Ramirez-Meyers, B. Rawn, and J. F. Whitacre, "A statistical assessment of the state-of-health of LiFePO4 cells harvested from a hybrid-electric vehicle battery pack," *J. Energy Storage*, vol. 59, p. 106472, Mar. 2023

[82]    F. Stroebl, R. Petersohn, B. Schricker, F. Schaeufl, O. Bohlen, and H. Palm, "A multi-stage lithium-ion battery aging dataset using various experimental design methodologies," *Sci. Data*, vol. 11, no. 1, p. 1020, Sept. 2024

[83]    K. Ramirez-Meyers, B. Rawn, and J. F. Whitacre, "A statistical assessment of the state-of-health of LiFePO4 cells harvested from a hybrid-electric vehicle battery pack," *J. Energy Storage*, vol. 59, p. 106472, Mar. 2023

[84]    E. Teliz, C. F. Zinola, and V. Díaz, "Identification and quantification of ageing mechanisms in li-ion batteries by electrochemical impedance spectroscopy.," *Electrochimica Acta*, vol. 426, p. 140801, Sept. 2022

[85]    Z. Li, J. Liu, Y. Qin, and T. Gao, "Enhancing the charging performance of lithium-ion batteries by reducing SEI and charge transfer resistances," *ACS Appl. Mater. Interfaces*, vol. 14, no. 29, pp. 33004–33012, July 2022

[86]    Z. Zeng *et al.*, "Physical interpretation of the electrochemical impedance spectroscopy (EIS) characteristics for diffusion-controlled intercalation and surface-redox charge storage behaviors," *J. Energy Storage*, vol. 102, p. 114021, Nov. 2024

[87]    Q. Zhang, C.-G. Huang, H. Li, G. Feng, and W. Peng, "Electrochemical impedance spectroscopy based state-of-health estimation for lithium-ion battery considering temperature and state-of-charge effect," *IEEE Trans. Transp. Electrification*, vol. 8, no. 4, pp. 4633–4645, Dec. 2022

[88]    Y. Yuan, B. Jiang, Q. Chen, X. Wang, X. Wei, and H. Dai, "A comparative study of battery state-of-charge estimation using electrochemical impedance spectroscopy by different machine learning methods," *Energy*, vol. 328, p. 136658, Aug. 2025

[89]    Y. Zhou, G. Dong, Q. Tan, X. Han, C. Chen, and J. Wei, "State of health estimation for lithium-ion batteries using geometric impedance spectrum features and recurrent gaussian process regression," *Energy*, vol. 262, p. 125514, Jan. 2023

[90]    M. Bao *et al.*, "Interpretable machine learning prediction for li-ion battery's state of health based on electrochemical impedance spectroscopy and temporal features," *Electrochimica Acta*, vol. 494, p. 144449, Aug. 2024

[91]    C. Li and L. Zhang, "Data-driven system identification, model transfer, and forecasting of battery capacity with impedance measurements," 2025, *SSRN*. Accessed: Nov. 28, 2025. [Online]. Available: https://www.ssrn.com/abstract=5173460

[92]    Z. Ning, P. Venugopal, T. Batista Soeiro, and G. Rietveld, "Computation-light AI models for robust battery capacity estimation based on electrochemical impedance spectroscopy," *IEEE Trans. Transp. Electrification*, vol. 11, no. 1, pp. 3146–3158, Feb. 2025

[93]    J. Zhu *et al.*, "Low-temperature separating lithium-ion battery interfacial polarization based on distribution of relaxation times (DRT) of impedance," *IEEE Trans. Transp. Electrification*, vol. 7, no. 2, pp. 410–421, June 2021

[94]    R. He, Y. He, W. Xie, B. Guo, and S. Yang, "Comparative analysis for commercial li-ion batteries degradation using the distribution of relaxation time method based on electrochemical impedance spectroscopy," *Energy*, vol. 263, p. 125972, Jan. 2023

[95]    J. W. Yap *et al.*, "Unveiling real-world aging mechanisms of lithium-ion batteries in electric vehicles," *J. Energy Storage*, vol. 130, p. 117420, Sept. 2025

[96]    F. Wang *et al.*, "SOH estimation for lithium-ion batteries using the distribution of relaxation time and feature optimized multilayer perceptron," *iScience*, vol. 28, no. 9, p. 113443, Sept. 2025

[97]    C.-Y. Yu *et al.*, "Time-resolved impedance spectroscopy analysis of aging in sulfide-based all-solid-state battery

full-cells using distribution of relaxation times technique," *J. Power Sources*, vol. 597, p. 234116, Mar. 2024

[98]  E. Braco, I. San Martin, P. Sanchis, A. Ursúa, and D.-I. Stroe, "Health indicator selection for state of health estimation of second-life lithium-ion batteries under extended ageing," *J. Energy Storage*, vol. 55, p. 105366, Nov. 2022

[99]  H. Feng and D. Song, "A health indicator extraction based on surface temperature for lithium-ion batteries remaining useful life prediction," *J. Energy Storage*, vol. 34, p. 102118, Feb. 2021

[100] P. Jindal, R. Katiyar, and J. Bhattacharya, "Evaluation of accuracy for bernardi equation in estimating heat generation rate for continuous and pulse-discharge protocols in LFP and NMC based li-ion batteries," *Appl. Therm. Eng.*, vol. 201, p. 117794, Jan. 2022

[101] J.-X. Li *et al.*, "Study on the temperature rise characteristics of aging lithium-ion batteries under different cooling methods," *Appl. Therm. Eng.*, vol. 240, p. 122235, Mar. 2024

[102] G. Zhang *et al.*, "Research on the impact of high-temperature aging on the thermal safety of lithium-ion batteries," *J. Energy Chem.*, vol. 87, pp. 378–389, Dec. 2023

[103] S. Chen, Y. Zhao, and S. Chen, "A consistency screening method for lithium-ion batteries based on infrared thermography and dynamic time warping," *J. Energy Storage*, vol. 132, p. 117922, Oct. 2025

[104] L. Giammichele, V. D'Alessandro, M. Falone, and R. Ricci, "Thermal behaviour assessment and electrical characterisation of a cylindrical lithium-ion battery using infrared thermography," *Appl. Therm. Eng.*, vol. 205, p. 117974, Mar. 2022

[105] Y. Yu *et al.*, "Distributed thermal monitoring of lithium ion batteries with optical fibre sensors," *J. Energy Storage*, vol. 39, p. 102560, July 2021

[106] Z. Wei, J. Hu, H. He, Y. Yu, and J. Marco, "Embedded distributed temperature sensing enabled multistate joint observation of smart lithium-ion battery," *IEEE Trans. Ind. Electron.*, vol. 70, no. 1, pp. 555–565, Jan. 2023

[107] L. Zhang, L. Liu, A. Terekhov, D. Warnberg, T. A. Zawodzinski, and P. Zhao, "The hotspot ignition nature of thermal runaway in li-ion batteries cycled under low temperatures," *J. Energy Storage*, vol. 129, p. 117272, Sept. 2025

[108] B. Hu *et al.*, "Early warning strategy for overheating-induced thermal runaway in lithium-ion batteries based on fast impedance measurement," *eTransportation*, vol. 26, p. 100498, Dec. 2025

[109] Y. Li *et al.*, "Battery eruption triggered by plated lithium on an anode during thermal runaway after fast charging," *Energy*, vol. 239, p. 122097, Jan. 2022

[110] Y. Zhang, S. Li, B. Mao, J. Shi, X. Zhang, and L. Zhou, "A multi-level early warning strategy for the LiFePO4 battery thermal runaway induced by overcharge," *Appl. Energy*, vol. 347, p. 121375, Oct. 2023

[111] Y. Li, S. Ding, L. Wang, W. Wang, C. Lin, and X. He, "On safety of swelled commercial lithium-ion batteries: A study on aging, swelling, and abuse tests," *eTransportation*, vol. 22, p. 100368, Dec. 2024

[112] M. Li *et al.*, "A safety warning method of li(Ni0.5Co0.2Mn0.3)O2 battery applicable for both overcharging and overheating conditions," *J. Energy Storage*, vol. 134, p. 117920, Oct. 2025

[113] S. Sun *et al.*, "Aging matters: How degradation pathways govern thermal runaway in lithium-ion batteries," *J. Energy Chem.*, vol. 114, pp. 10–21, Mar. 2026

[114] G. Zhang *et al.*, "Comprehensive investigation of a slight overcharge on degradation and thermal runaway behavior of lithium-ion batteries," *ACS Appl. Mater. Interfaces*, vol. 13, no. 29, pp. 35054–35068, July 2021

[115] E. Braco, I. San Martín, A. Berrueta, P. Sanchis, and A. Ursúa, "Experimental assessment of cycling ageing of lithium-ion second-life batteries from electric vehicles," *J. Energy Storage*, vol. 32, p. 101695, Dec. 2020

[116] M. Luh and T. Blank, "Comprehensive battery aging dataset: Capacity and impedance fade measurements of a lithium-ion NMC/C-SiO cell," *Sci. Data*, vol. 11, no. 1, p. 1004, Sept. 2024

[117] Z. Zhou *et al.*, "An efficient screening method for retired lithium-ion batteries based on support vector machine," *J. Clean. Prod.*, vol. 267, p. 121882, Sept. 2020

[118] P. Huang, Y. Zhang, Y. Kang, P. Gu, B. Duan, and C. Zhang, "A flexible screening scheme for retired lithium-ion batteries based on novel capacity indicator and random forest algorithm," *IEEE Trans. Transp. Electrification*, vol. 11, no. 1, pp. 544–557, Feb. 2025

[119] J. He, X. Bian, L. Liu, Z. Wei, and F. Yan, "Comparative study of curve determination methods for incremental capacity analysis and state of health estimation of lithium-ion battery," *J. Energy Storage*, vol. 29, p. 101400, June 2020

[120] J. Zhu *et al.*, "Investigation of lithium-ion battery degradation mechanisms by combining differential voltage analysis and alternating current impedance," *J. Power Sources*, vol. 448, p. 227575, Feb. 2020

[121] A. G. Li, A. C. West, and M. Preindl, "Characterizing degradation in lithium-ion batteries with pulsing," *J. Power Sources*, vol. 580, p. 233328, Oct. 2023

[122] J. Sun and J. Kainz, "Optimization of hybrid pulse power characterization profile for equivalent circuit model parameter identification of li-ion battery based on taguchi method," *J. Energy Storage*, vol. 70, p. 108034, Oct. 2023

[123] A. G. Li, M. Berecibar, and M. Preindl, "Nonlinear characterization of lithium-ion batteries with bipolar pulsing," *IEEE Trans. Ind. Electron.*, vol. 71, no. 10, pp. 12983–12990, Oct. 2024

[124] R. Wang *et al.*, "Degradation analysis of lithium-ion batteries under ultrahigh-rate discharge profile," *Appl. Energy*, vol. 376, p. 124241, Dec. 2024

[125] C. Li *et al.*, "SOH estimation method for lithium-ion batteries based on an improved equivalent circuit model via electrochemical impedance spectroscopy," *J. Energy Storage*, vol. 86, p. 111167, May 2024

[126] S. Cruz-Manzo and P. Greenwood, "An impedance model based on a transmission line circuit and a frequency dispersion warburg component for the study of EIS in li-ion batteries," *J. Electroanal. Chem.*, vol. 871, p. 114305, Aug. 2020

[127] Y. Zhang, Q. Tang, Y. Zhang, J. Wang, U. Stimming, and A. A. Lee, "Identifying degradation patterns of lithium ion batteries from impedance spectroscopy using machine learning," *Nat. Commun.*, vol. 11, no. 1, p. 1706, Apr. 2020

[128] F. Luo, H. Huang, L. Ni, and T. Li, "Rapid prediction of the state of health of retired power batteries based on electrochemical impedance spectroscopy," *J. Energy Storage*, vol. 41, p. 102866, Sept. 2021

[129] J. Tian, R. Xiong, and W. Shen, "State-of-Health Estimation Based on Differential Temperature for Lithium Ion Batteries," *IEEE Trans. Power Electron.*, vol. 35, no. 10, pp. 10363–10373, Oct. 2020

[130] C. Wang *et al.*, "Thermal runaway behavior and features of LiFePO$_4$/graphite aged batteries under overcharge," *Int. J. Energy Res.*, vol. 44, no. 7, pp. 5477–5487, June 2020

[131] K. Li *et al.*, "Early warning for thermal runaway in lithium-ion batteries during various charging rates: Insights from expansion force analysis," *J. Clean. Prod.*, vol. 457, p. 142422, June 2024

[132] T. Jia, Y. Zhang, C. Ma, S. Li, H. Yu, and G. Liu, "The early warning for overcharge thermal runaway of lithium-ion batteries based on a composite parameter," *J. Power Sources*, vol. 555, p. 232393, Jan. 2023

[133] Z. Wei *et al.*, "Machine learning-based hybrid thermal modeling and diagnostic for lithium-ion battery enabled by embedded sensing," *Appl. Therm. Eng.*, vol. 216, p. 119059, Nov. 2022

[134] A. Legala and X. Li, "Hybrid data-based modeling for the prediction and diagnostics of li-ion battery thermal behaviors," *Energy AI*, vol. 10, p. 100194, Nov. 2022

[135] Z. Zhou, B. Duan, Y. Kang, Y. Shang, Q. Zhang, and C. Zhang, "Health feature extraction and efficient sorting of second-life lithium-ion batteries," *IEEE Trans. Energy Convers.*, vol. 39, no. 2, pp. 1373–1382, June 2024

[136] S. Zhou, W. Du, B. Mager, P. R. Shearing, T. S. Miller, and R. Jervis, "Batch diagnosis of batteries within one second," *Joule*, vol. 9, no. 9, p. 102099, Sept. 2025

[137] J. Zhao, X. Feng, M.-K. Tran, M. Fowler, M. Ouyang, and A. F. Burke, "Battery safety: Fault diagnosis from laboratory to real world," *J. Power Sources*, vol. 598, p. 234111, Apr. 2024

[138] M. Fichtner *et al.*, "Rechargeable batteries of the future—the state of the art from a BATTERY 2030+ perspective," *Adv. Energy Mater.*, vol. 12, no. 17, p. 2102904, May 2022

[139] S. Tao *et al.*, "Collaborative and privacy-preserving retired battery sorting for profitable direct recycling via federated machine learning," *Nat. Commun.*, vol. 14, no. 1, p. 8032, Dec. 2023

[140] M. Zhao, Y. Zhang, and H. Wang, "Battery degradation stage detection and life prediction without accessing historical operating data," *Energy Storage Mater.*, vol. 69, p. 103441, May 2024

[141] V. Sulzer *et al.*, "The challenge and opportunity of battery lifetime prediction from field data," *Joule*, vol. 5, no. 8, pp. 1934–1955, Aug. 2021

[142] S. Tao, "Generative learning assisted state-of-health estimation for sustainable battery recycling with random retirement conditions," *Nat. Commun.*, 2024

[143] W. Li, J. Chen, K. Quade, D. Luder, J. Gong, and D. U. Sauer, "Battery degradation diagnosis with field data, impedance-based modeling and artificial intelligence," *Energy Storage Mater.*, vol. 53, pp. 391–403, Dec. 2022

[144] Y. Zhang, J. Xie, C. Wang, Q. Xie, and N. Wang, "State-of-health estimation of lithium-ion batteries based on self-transformer with few-labeled data and label noise," *J. Power Sources*, vol. 649, p. 237424, Sept. 2025

[145] T. Wang, Z. Ma, S. Zou, Z. Chen, and P. Wang, "Lithium-ion battery state-of-health estimation: A self-supervised framework incorporating weak labels," *Appl. Energy*, vol. 355, p. 122332, Feb. 2024

[146] Y. Xie *et al.*, "Inhomogeneous degradation induced by lithium plating in a large-format lithium-ion battery," *J. Power Sources*, vol. 542, p. 231753, Sept. 2022

[147] L. Wang *et al.*, "A novel OCV curve reconstruction and update method of lithium-ion batteries at different temperatures based on cloud data," *Energy*, vol. 268, p. 126773, Apr. 2023

[148] X. Xu, Z. Xu, T. Wang, J. Xu, and L. Pei, "Open-circuit voltage curve reconstruction for degrading lithium-ion batteries utilizing discrete curve fragments from an online dataset," *J. Energy Storage*, vol. 56, p. 106003, Dec. 2022

[149] D. Doonyapisut, B. Kim, J. K. Kim, E. Lee, and C.-H. Chung, "Deep generative learning for exploration in large electrochemical impedance dataset," *Eng. Appl. Artif. Intell.*, vol. 126, p. 107027, Nov. 2023

[150] J. Li, X. Li, X. Yuan, and Y. Zhang, "Deep learning method for online parameter identification of lithium-ion batteries using electrochemical synthetic data," *Energy Storage Mater.*, vol. 72, p. 103697, Sept. 2024

[151] H. Jin *et al.*, "A synthetic data generation method and evolutionary transformer model for degradation trajectory prediction in lithium-ion batteries," *Appl. Energy*, vol. 377, p. 124629, Jan. 2025

[152] M. Duquesnoy, C. Liu, D. Z. Dominguez, V. Kumar, E. Ayerbe, and A. A. Franco, "Machine learning-assisted multi-objective optimization of battery manufacturing from synthetic data generated by physics-based simulations," *Energy Storage Mater.*, vol. 56, pp. 50–61, Feb. 2023

[153] J. Kim *et al.*, "A physics-driven generative model to accelerate artificial intelligence development for lithium-ion battery diagnostics," *Appl. Energy*, vol. 391, p. 125873, Aug. 2025

[154] W. Yao, R. Lai, Y. Tian, X. Li, and J. Tian, "State of Health Estimation of Lithium-Ion Batteries Using Data Augmentation and Feature Mapping," *IEEE Trans. Transp. Electrification*, vol. 11, no. 1, pp. 4895–4905, Feb. 2025

[155] S. Echabarri, P. Do, H.-C. Vu, and P.-Y. Liegeois, "A modified TimeGAN-based data augmentation approach for the state of health prediction of lithium-ion batteries," *Reliab. Eng. Syst. Saf.*, vol. 264, p. 111297, Dec. 2025

[156] J. Sun, A. Gu, and J. Kainz, "A solution framework for the experimental data shortage problem of lithium-ion batteries: Generative adversarial network-based data augmentation for battery state estimation," *J. Energy Chem.*, vol. 103, pp. 476–497, Apr. 2025

[157] X. Liu, Z. Gao, J. Tian, Z. Wei, C. Fang, and P. Wang, "State of Health Estimation for Lithium-ion Batteries Using Voltage Curves Reconstruction by Conditional Generative Adversarial Network," *IEEE Trans. Transp. Electrification*, pp. 1–1, 2024

[158] H. Xiang, Y. Wang, Y.-Y. Soo, X. Xiong, and Z. Chen, "Predicting Automotive Battery Degradation Trajectories using Transformer with Variational Auto-encoder based Data Augmentation," *IEEE Trans. Veh. Technol.*, pp. 1–12, 2025

[159] Y. Huang, Y. Tang, and J. VanZwieten, "Prognostics With Variational Autoencoder by Generative Adversarial Learning," *IEEE Trans. Ind. Electron.*, vol. 69, no. 1, pp. 856–867, Jan. 2022

[160] L. Zhang, B. Chen, D. Lyu, L. Wang, and K. Wang, "Estimation of lithium-ion battery health state by evae and bigru based on eis," 2025, *SSRN*. Accessed: Nov. 30, 2025. [Online]. Available: https://www.ssrn.com/abstract=5100808

[161] Z. Yang, Y. Pan, W. Liu, J. Meng, and Z. Song, "Enhanced fault detection in lithium-ion battery energy storage systems via transfer learning-based conditional GAN under limited data," *J. Power Sources*, vol. 645, p. 237192, July 2025

[162] X. Lu, J. Qiu, G. Lei, and J. Zhu, "State of health estimation of lithium iron phosphate batteries based on degradation knowledge transfer learning," *IEEE Trans. Transp. Electrification*, vol. 9, no. 3, pp. 4692–4703, Sept. 2023

[163] P. J. Weddle *et al.*, "Battery state-of-health diagnostics during fast cycling using physics-informed deep-learning," *J. Power Sources*, vol. 585, p. 233582, Nov. 2023

[164] L. Yang, M. He, Y. Ren, B. Gao, and H. Qi, "Physics-informed neural network for co-estimation of state of health, remaining useful life, and short-term degradation path in lithium-ion batteries," *Appl. Energy*, vol. 398, p. 126427, Nov. 2025

[165] P. Wen, Z.-S. Ye, Y. Li, S. Chen, P. Xie, and S. Zhao, "Physics-informed neural networks for prognostics and health management of lithium-ion batteries," *IEEE Trans. Intell. Veh.*, vol. 9, no. 1, pp. 2276–2289, Jan. 2024

[166] S. Zhang, Z. Liu, Y. Xu, J. Guo, and H. Su, "A physics-informed hybrid data-driven approach with generative electrode-level features for lithium-ion battery health prognostics," *IEEE Trans. Transp. Electrification*, vol. 11, no. 1, pp. 4857–4871, Feb. 2025

[167] G. Sun, Y. Liu, and X. Liu, "A method for estimating lithium-ion battery state of health based on physics-informed machine learning," *J. Power Sources*, vol. 627, p. 235767, Jan. 2025

[168] J. Shi, A. Rivera, and D. Wu, "Battery health management using physics-informed machine learning: Online degradation modeling and remaining useful life prediction," *Mech. Syst. Signal Process.*, vol. 179, p. 109347, Nov. 2022

[169] W. Deng *et al.*, "A generic physics-informed machine learning framework for battery remaining useful life prediction using small early-stage lifecycle data," *Appl. Energy*, vol. 384, p. 125314, Apr. 2025

[170] X. Nie, Y. Pan, Y. Zhang, Z. Luo, and S. Wang, "Battery health prediction under data scarcity: A cross-domain physics-informed 5-shot framework with GRU-transformer," *Appl. Energy*, vol. 402, p. 127012, Jan. 2026

[171] S. Zhang, M. Liu, R. Guo, J. Tian, Z. Man, and W. Shen, "Battery life prediction with scarce data using physics-informed data generation and adaptive autoencoder," *IEEE Trans. Transp. Electrification*, pp. 1–1, 2025

[172] X. Li, D. Yu, S. B. Vilsen, and D. I. Stroe, "Accuracy comparison and improvement for state of health estimation of lithium-ion battery based on random partial recharges and feature engineering," *J. Energy Chem.*, vol. 92, pp. 591–604, May 2024

[173] L. Cai, J. Meng, D.-I. Stroe, J. Peng, G. Luo, and R. Teodorescu, "Multiobjective optimization of data-driven model for lithium-ion battery SOH estimation with short-term feature," *IEEE Trans. Power Electron.*, vol. 35, no. 11, pp. 11855–11864, Nov. 2020

[174] N. Li, F. He, W. Ma, R. Wang, L. Jiang, and X. Zhang, "An indirect state-of-health estimation method based on improved genetic and back propagation for online lithium-ion battery used in electric vehicles," *IEEE Trans. Veh. Technol.*, vol. 71, no. 12, pp. 12682–12690, Dec. 2022

[175] S. Li, P. Zhao, C. Gu, D. Huo, J. Li, and S. Cheng, "Linearizing battery degradation for health-aware vehicle

energy management," *IEEE Trans. Power Syst.*, vol. 38, no. 5, pp. 4890–4899, Sept. 2023

[176] S. Park, J. Kim, and I. Cho, "Hybrid SOC and SOH estimation method with improved noise immunity and computational efficiency in hybrid railroad propulsion system," *J. Energy Storage*, vol. 72, p. 108385, Nov. 2023

[177] K. S. Mayilvahanan, K. J. Takeuchi, E. S. Takeuchi, A. C. Marschilok, and A. C. West, "Supervised learning of synthetic big data for li-ion battery degradation diagnosis," *Batter. Supercaps*, vol. 5, no. 1, p. e202100166, Jan. 2022

[178] Z. Ye, J. Chang, and J. Yu, "Prognosability regularized generative adversarial network for battery state of health estimation with limited samples," *Energy*, vol. 325, p. 135922, June 2025

[179] H. Wang and Y.-F. Li, "Robust mechanical fault diagnosis with noisy label based on multistage true label distribution learning," *IEEE Trans. Reliab.*, vol. 72, no. 3, pp. 975–988, Sept. 2023

[180] Y. Xie, L. Dang, H. Cao, and B. Chen, "A robust semi-supervised confidence-aware framework for battery state of health estimation," *J. Energy Storage*, vol. 139, p. 118842, Dec. 2025

[181] S. Zhang, Y. Li, J. Tian, Z. Man, C. Y. Chung, and W. Shen, "Improving Battery Life Prediction With Unlabeled Data: Confidence-Weighted Semi-Supervised Learning With Label Propagation," *IEEE Trans. Transp. Electrification*, vol. 11, no. 2, pp. 5938–5949, Apr. 2025

[182] Y. Cao, M. Jia, X. Zhao, X. Yan, and Z. Liu, "Semi-supervised machinery health assessment framework via temporal broad learning system embedding manifold regularization with unlabeled data," *Expert Syst. Appl.*, vol. 222, p. 119824, July 2023

[183] Y.-X. Wang, S. Zhao, S. Wang, K. Ou, and J. Zhang, "Enhanced vision-transformer integrating with semi-supervised transfer learning for state of health and remaining useful life estimation of lithium-ion batteries," *Energy AI*, vol. 17, p. 100405, Sept. 2024

[184] Y. Xiang, W. Fan, J. Zhu, X. Wei, and H. Dai, "Semi-supervised deep learning for lithium-ion battery state-of-health estimation using dynamic discharge profiles," *Cell Rep. Phys. Sci.*, vol. 5, no. 1, p. 101763, Jan. 2024

[185] Y. Liu, J. Ding, Y. Cai, B. Luo, L. Yao, and Z. Wang, "A battery SOH estimation method based on entropy domain features and semi-supervised learning under limited sample conditions," *J. Energy Storage*, vol. 106, p. 114822, Jan. 2025

[186] F. Naaz, A. Herle, J. Channegowda, A. Raj, and M. Lakshminarayanan, "A generative adversarial network-based synthetic data augmentation technique for battery condition evaluation," *Int. J. Energy Res.*, vol. 45, no. 13, pp. 19120–19135, Oct. 2021

[187] S. Wang, J. Li, G. Hou, and D. Yuan, "Application of online semi-supervised learning embedded with chaotic dynamics in equipment health prognostics," *IEEE Access*, vol. 13, pp. 103982–103994, 2025

[188] T. Wang, R. Chao, Z. Dong, and L. Feng, "Weakly supervised battery SOH estimation with imprecise intervals," *IEEE Trans. Energy Convers.*, vol. 40, no. 3, pp. 1841–1855, Sept. 2025

[189] H. Xiang, Y. Wang, X. Zhang, and Z. Chen, "Two-level battery health diagnosis using encoder–decoder framework and gaussian mixture ensemble learning based on relaxation voltage," *IEEE Trans. Transp. Electrification*, vol. 10, no. 2, pp. 3966–3975, June 2024

[190] W. Guo, L. Yang, Z. Deng, B. Xiao, and X. Bian, "Early diagnosis of battery faults through an unsupervised health scoring method for real-world applications," *IEEE Trans. Transp. Electrification*, vol. 10, no. 2, pp. 2521–2532, June 2024

[191] M. S. Batta, H. Mabed, Z. Aliouat, and S. Harous, "Battery state-of-health prediction-based clustering for lifetime optimization in IoT networks," *IEEE Internet Things J.*, vol. 10, no. 1, pp. 81–91, Jan. 2023

[192] J. Hu, B. Lin, M. Wang, J. Zhang, W. Zhang, and Y. Lu, "Health factor analysis and remaining useful life prediction for batteries based on a cross-cycle health factor clustering framework," *J. Energy Storage*, vol. 50, p. 104661, June 2022

[193] X. Qiu, Y. Bai, and S. Wang, "A novel unsupervised domain adaptation-based method for lithium-ion batteries

state of health prognostic," *J. Energy Storage*, vol. 75, p. 109358, Jan. 2024

[194]   F. Li, Y. Yu, X. Yuan, and G. Ren, "State-of-health estimation for lithium-ion batteries using unsupervised deep subdomain adaptation," *Energy*, vol. 324, p. 135862, June 2025

[195]   Z. Ye and J. Yu, "State-of-health estimation for lithium-ion batteries using domain adversarial transfer learning," *IEEE Trans. Power Electron.*, vol. 37, no. 3, pp. 3528–3543, Mar. 2022

[196]   M. Badfar, M. Yildirim, and R. B. Chinnam, "State-of-charge estimation across battery chemistries: A novel regression-based method and insights from unsupervised domain adaptation," *J. Power Sources*, vol. 628, p. 235760, Feb. 2025

[197]   Y. Che, Y. Zheng, X. Sui, and R. Teodorescu, "Boosting battery state of health estimation based on self-supervised learning," *J. Energy Chem.*, vol. 84, pp. 335–346, Sept. 2023

[198]   F. Heinrich, F. K.-D. Noering, M. Pruckner, and K. Jonas, "Unsupervised data-preprocessing for long short-term memory based battery model under electric vehicle operation," *J. Energy Storage*, vol. 38, p. 102598, June 2021